\documentclass[aps,pre,groupedaddress,showpacs,preprint]{revtex4}
\usepackage{graphicx}
\usepackage[dvips]{epsfig}
\usepackage{dcolumn}
\usepackage{subfigure}%
\usepackage{bm}
\usepackage{amssymb}
\usepackage{amsmath}

\begin{document}
\title{Statistics of interfacial fluctuations of radially growing clusters}

\author{Carlos Escudero}

\affiliation{Departamento de Econom\'{\i}a Cuantitativa \& \\
Instituto de Ciencias Matem\'aticas (CSIC-UAM-UC3M-UCM), \\
Universidad Aut\'onoma de Madrid, \\
Ciudad Universitaria de Cantoblanco, 28049 Madrid, Spain}

\begin{abstract}
The dynamics of fluctuating radially growing interfaces is
approached using the formalism of stochastic growth equations on
growing domains. This framework reveals a number of dynamic
features arising during surface growth. For fast growth, dilution,
which spatially reorders the incoming matter, is responsible for
the transmission of correlations. Its effects include the erasing
of memory with respect to the initial condition, a partial
attenuation of geometrically originated instabilities, and the
restoring of universality in some special cases in which the
critical exponents depend on the parameters of the equation of
motion. In this sense, dilution rends the dynamics more similar to
the usual one of planar systems. This fast growth regime is also
characterized by the spatial decorrelation of the interface, which
in the case of radially growing interfaces naturally originates
rapid roughening and scale dependent fractality, and suggests the
advent of a self-similar fractal dimension. The center of mass
fluctuations of growing clusters are also studied, and our
analysis suggests the possible non-applicability of usual scalings
to the long range surface fluctuations of the radial Eden model.
In fact, our study points to the fact that this model belongs to a
dilution-free universality class.
\end{abstract}


\maketitle

\section{Introduction}

The study of fluctuating interfaces has occupied an important
place within statistical mechanics in recent and not so recent
times. The origins of this interest are practical, due to the vast
range of potential applications that this theory may have, and
theoretical, as some of the universality classes discovered within
this framework are claimed to play an important role in other
areas of physics~\cite{barabasi}. While the great majority of
works on this topic has concentrated on strip or slab geometries,
it is true that at the very beginning of the theoretical studies
on nonequilibrium growth one finds the seminal works by Eden,
focused on radial shapes~\cite{eden1,eden2}. To a certain extent,
the motivation of considering radial forms is related to
biological growth, as for instance the Eden model can be thought
of as a simplified description of a developing cell colony. The
Eden and other related discrete models have been computationally
analyzed along the years, and the results obtained have been put
in the context of stochastic growth theory, see for
instance~\cite{ferreira} and references therein. One of the most
important models of nonequilibrium growth of radial systems is
diffusion-limited aggregation (DLA)~\cite{sander}. It is related
to physical phenomena such as electrodeposition, Hele-Shaw flow,
mineral deposits and dielectric breakdown. The theory we will
present in this paper is not in a state to be able to describe
such complex branched structures. Nevertheless, as we will see in
the following, it might be related to specific limits of DLA
processes.

Apart from the interest in modelling, there is a genuine
theoretical motivation in understanding the dynamics of growing
radial clusters. The Eden model is actually a sort of first
passage percolation~\cite{hammersley}, and the scaling limit of
percolation models has been studied by means of field-theoretic
approaches~\cite{hinrichsen} and stochastic processes like
Schramm-Loewner evolution~\cite{sle}. A natural theoretical
question to be answered is in which cases the Family-Vicsek
scaling~\cite{family}, basic to describe planar growth processes,
is able to capture the behavior of the surface fluctuations of
growing radial clusters.

The use of stochastic differential equations, very much spread in
the modelling of planar growth profiles, has been not so commonly
employed in the case of radial growth. A series of works
constitute an exception to this rule
\cite{kapral,batchelor,singha,cescudero1,cescudero2,escudero,escudero2},
as they proposed a partial differential equation with stochastic
terms as a benchmark for analyzing the dynamics of radial
interfaces. Because studying this sort of equations is complicated
by the nonlinearities implied by reparametrization invariance, a
simplified version in which only the substrate growth was
considered was introduced in \cite{escuderojs}. Already in this
case it was apparent that for rapidly growing interfaces dilution,
which is responsible for matter redistribution as the substrate
grows~\cite{maini}, propagates the correlations when large
spatiotemporal scales are considered. It is also capable of
erasing the memory effects that would otherwise arise, let us show
how. In \cite{escuderojs} we considered the linear equation for
stochastic growth on a growing domain
\begin{equation}
\label{gdomain}
\partial_t h=-D \left( \frac{t_0}{t} \right)^{\zeta \gamma} |\nabla|^\zeta h
-\frac{d\gamma}{t}h +\gamma F t^{\gamma-1}+
\left(\frac{t_0}{t}\right)^{d\gamma/2}\xi(x,t),
\end{equation}
where the domain grows following the power law $t^\gamma$,
$\gamma>0$ is the growth index and $-(d\gamma/t)h$ is the term
taking into account dilution~\cite{escuderojs}. Dilution refers to
the fact that the interfacial matter, as the interface grows,
becomes redistributed in a larger domain. Its Fourier transformed
version, for $n \ge 1$, is
\begin{equation}
\frac{d h_n}{dt}=-D \left( \frac{t_0}{t} \right)^{\zeta \gamma}
\frac{\pi^\zeta |n|^\zeta}{L_0^\zeta} h_n -\frac{d\gamma}{t}h_n +
\left(\frac{t_0}{t}\right)^{d\gamma/2}\xi_n(t).
\end{equation}
This equation can be readily solved for $\gamma > 1/\zeta$ and in
the long time limit
\begin{equation}
h_n(t)=(t/t_0)^{-d \gamma} \exp \left[ \frac{D t_0}{1-\zeta
\gamma} \frac{\pi^\zeta |n|^\zeta}{L_0^\zeta} \right] h_n(t_0)+
(t/t_0)^{-d \gamma} \int_{t_0}^t \left( \frac{\tau}{t_0}
\right)^{d\gamma/2}\xi_n(\tau) d\tau,
\end{equation}
and so the dependence on the initial condition tends to zero as a
power law for long times. This is, as mentioned, one of the
consequences of dilution. If we considered the dilatation
transformation $x \to (t/t_0)^\gamma x$ we would find again
Eq.~(\ref{gdomain}) but this time without the dilution term. This
corresponds to a dilatation of mass and space simultaneously. The
solution now becomes
\begin{equation}
h_n(t)= \exp \left[ \frac{D t_0}{1-\zeta \gamma} \frac{\pi^\zeta
|n|^\zeta}{L_0^\zeta} \right] h_n(t_0)+ \int_{t_0}^t
\left(\frac{t_0}{\tau}\right)^{d\gamma/2}\xi_n(\tau)d\tau,
\end{equation}
and so the dependence on the initial condition remains for all
times. In the first case the long time solution becomes spatially
uncorrelated, and in the second one only part of the initial
correlations survive. As an abuse of language, we will talk about
decorrelation in both cases. The memory effects that affect the
solution in the no-dilution (or dilatation) situation separate its
behavior from the one dictated by the Family-Vicsek scaling
\cite{escuderojs,escuderoar}. For $\gamma < 1/\zeta$ the memory
effects and the corresponding dependence on the initial condition
disappear exponentially fast for long times as a consequence of
the effect of diffusion.

We start discussing dilution as the mechanism that controls the
amount of matter on the interface. Pure diffusion on a growing
domain is described by the equation
\begin{equation}
\partial_t h= D \left( \frac{t_0}{t} \right)^{2 \gamma} \nabla^2 h -\frac{d \gamma}{t}h,
\end{equation}
in Eulerian coordinates $x \in [0,L_0] \times \cdots \times
[0,L_0]$ (see~\cite{escuderojs}) and where dilution has been taken
into account. The total mass on the surface is conserved
\begin{equation}
\int_0^{L(t)} \cdots \int_0^{L(t)}h(y,t)dy= \left( \frac{t}{t_0}
\right)^{d \gamma} \int_0^{L_0} \cdots \int_0^{L_0} h(x,t)dx=
\int_0^{L_0} \cdots \int_0^{L_0}h(x,t_0)dx,
\end{equation}
where $y \equiv [L(t)/L_0]x$ denotes the set of Lagrangian
coordinates. Note we are using the shorthand notations
$x=(x_1,\cdots,x_d)$, $y=(y_1, \cdots,y_d)$, $dx=dx_1 \cdots
dx_d$, and $dy=dy_1 \cdots dy_d$. In the no-dilution situation we
find
\begin{equation}
\int_0^{L(t)} \cdots \int_0^{L(t)}h(y,t)dy= \left( \frac{t}{t_0}
\right)^{d \gamma} \int_0^{L_0} \cdots \int_0^{L_0} h(x,t)dx=
\left( \frac{t}{t_0} \right)^{d \gamma} \int_0^{L_0} \cdots
\int_0^{L_0}h(x,t_0)dx.
\end{equation}
This second case is pure dilatation, which implies that not only
the space grows, but also the interfacial matter grows at the same
rate, in such a way that the average density remains constant.
Note that this process of matter dilatation, as well as the
spatial growth, are deterministic processes. These calculations
show that both dilution and dilatation dynamics are physically
motivated and have a number of measurable differences. It is worth
remarking here that all previous works
except~\cite{escuderojs,escuderolast} have exclusively considered
dilatation dynamics. Even in the different field of
reaction-diffusion dynamics in which the dilution term was
derived, the focus was on the limit in which it was
irrelevant~\cite{maini}.

This work is devoted to further explore the consequences of
dilution, dilatation and decorrelation, and their effects on
scaling of radial interfaces. We will use in cases radial
stochastic growth equations, which may show up
instabilities~\cite{escudero2}, and explore the interplay of
dilution with them. In other cases, when instabilities do not play
a determinant role and for the sake of simplicity, we will
consider stochastic growth equations on growing domains. The
outlook of the paper is as follow: In Sec.~\ref{rrd} we consider
the simplest radial growth process, radial random deposition, and
derive for the first time the two-point space-time correlation
functions. In Sec.~\ref{rdad} and Sec.~\ref{instabilities} we
compute for the first time the two-point space-time correlators
for stochastic growth equations taking into account simultaneously
random deposition and diffusion in the absence and presence of
instabilities respectively. In Sec.~\ref{rapidrough} and
Sec.~\ref{polyfractal} we show, for the first time, that radial
stochastic growth equations give rise naturally to the phenomena
of rapid roughening and scale dependent fractality of surfaces. In
Sec.~\ref{kpz} we mention some of the problems that arise when
studying the radial counterpart of nonlinear stochastic growth
equations and in Sec.~\ref{comfluct} we calculate for the first
time the center of mass fluctuations of the cluster interfaces
described by radial stochastic growth equations. Finally, in
Sec.~\ref{eden} we apply our results to Eden clusters and in
Sec.~\ref{outlook} we draw our main conclusions.

\section{Radial Random Deposition}
\label{rrd}

In this section we construct for the first time two-point
space-time correlation functions for the radial random deposition
process. This quantities are fundamental in order to statistically
characterize fluctuating interfaces.

In order to construct radial growth equations one may invoke the
reparametrization invariance principle~\cite{maritan1,maritan2},
as has already been done a number of
times~\cite{kapral,batchelor,cescudero1,cescudero2,escudero,escudero2}.
In case of white and Gaussian fluctuations, the $d-$dimensional
spherical noise is given by
\begin{eqnarray}
\frac{1}{\sqrt[4]{g \left[\vec{\theta},r(\vec{\theta},t)
\right]}}\xi(\vec{\theta},t),
\qquad \left< \xi(\vec{\theta},t) \right>=0, \\
\left< \xi(\vec{\theta},t) \xi(\vec{\theta},t) \right>= \epsilon
\delta(\vec{\theta}-\vec{\theta}') \delta(t-t'),
\end{eqnarray}
where $g=\det(g_{ij})= \det(\partial_i \vec{r} \cdot \partial_j
\vec{r})$ is the determinant of the metric tensor. Under the small
gradient assumption $|\nabla_{\vec{\theta}} \,\, r| \ll r$ one
finds $g \approx \mathcal{J}(r,\vec{\theta})^2$, where
$\mathcal{J}$ is the Jacobian determinant of the change of
variables from the Cartesian representation $(\vec{x},h)$ to the
polar representation $(\vec{\theta},r)$. We also have the
factorization $\mathcal{J}(r,\vec{\theta})^2= r^{2d}
J(\vec{\theta})^2$, where $J$ is the Jacobian evaluated at $r=1$.

The simplest growth process is possibly the radial random
deposition model. If the growth rate is explicitly time dependent,
then the growth equation reads
\begin{equation}
\label{rdeposition}
\partial_t r = F \gamma
t^{\gamma-1}+\frac{1}{r^{d/2}J(\vec{\theta})^{1/2}}\xi(\vec{\theta},t),
\end{equation}
in the absence of dilution. Here $r(\vec{\theta},t)$ is the value
of the radius at the angular position $\vec{\theta}$ and time $t$,
$F>0$ is the constant prefactor of the growth rate, $\gamma>0$ is
the growth index, $d$ is the number of angles used to parameterize
the cluster surface (so the cluster grows in $d+1$ spatial
dimensions) and $\xi$ is a zero mean Gaussian noise, whose
correlation is given by
\begin{equation}
\left< \xi(\vec{\theta},t) \xi(\vec{\theta}',s) \right>= \epsilon
\delta(\vec{\theta}-\vec{\theta}') \delta(t-s).
\end{equation}
The equation for the first moment can be easily obtained
\begin{equation}
\partial_t \left< r \right> = F \gamma t^{\gamma-1},
\end{equation}
due to the It\^o interpretation of the noise, and we can integrate
it to get
\begin{equation}
\left< r(\vec{\theta},t) \right> = F t^{\gamma},
\end{equation}
where we have assumed the radially symmetric initial condition
$r(\vec{\theta},t_0)=F t_0^{\gamma}$ and $t_0 \le t$ is the
absolute origin of time. It is difficult to obtain more
information from the full equation~(\ref{rdeposition}), so we will
perform a perturbative expansion. We assume the solution form
\begin{equation}
\label{snoise} r(\vec{\theta},t)= R(t)+
\sqrt{\epsilon}\rho_1(\vec{\theta},t),
\end{equation}
where the noise intensity $\epsilon$ will be used as the small
parameter~\cite{gardiner}. Substituting this solution form into
Eq.~(\ref{rdeposition}) we obtain the equations
\begin{eqnarray}
\partial_t R &=& F \gamma t^{\gamma-1}, \\
\partial_t \rho_1 &=& \frac{1}{F^{d/2}t^{\gamma d/2}} \frac{\eta(\vec{\theta},t)}{J(\vec{\theta})^{1/2}},
\end{eqnarray}
where $\xi=\sqrt{\epsilon} \, \eta$. These equations have been
derived assuming $\sqrt{\epsilon} \ll F t^\gamma$, a condition
much more favorable (the better the larger $\gamma$ is) than the
usual time independent ones supporting small noise
expansions~\cite{gardiner}. The solution to these equations can be
readily computed
\begin{eqnarray}
R(\vec{\theta},t) &=& Ft^\gamma, \\
\left< \rho_1(\vec{\theta},t) \right> &=& 0, \\
\left< \rho_1(\vec{\theta},t) \rho_1(\vec{\theta}',s) \right> &=&
\frac{F^{-d}}{1-\gamma d} \left[ \left( \min \{t,s\}
\right)^{1-\gamma d}-t_0^{1-\gamma d}\right]
\frac{\delta(\vec{\theta}-\vec{\theta}')}{J(\vec{\theta})},
\end{eqnarray}
if $\gamma d \neq 1$ and where we have assumed a zero value for
the initial perturbation. If $\gamma d = 1$ the correlation
becomes
\begin{equation}
\left< \rho_1(\vec{\theta},t) \rho_1(\vec{\theta}',s) \right> =
\frac{1}{F^{d}} \mathrm{ln}\left[ \frac{\min \{t,s\}}{t_0}
\right]\frac{\delta(\vec{\theta}-\vec{\theta}')}{J(\vec{\theta})}.
\end{equation}
Here $R$ is a deterministic function and $\rho_1$ is a zero mean
Gaussian stochastic process that is completely determined by the
correlations given above. The long time behavior of the
correlations, given by the condition $t,s \gg t_0$, is specified
by the following two-times and one-time correlation functions
\begin{eqnarray}
\left< \rho_1(\vec{\theta},t) \rho_1(\vec{\theta}',s) \right> &=&
\frac{F^{-d}}{1-\gamma d}\left( \min\{ t,s \} \right)^{1-\gamma d}
\frac{\delta(\vec{\theta}-\vec{\theta}')}{J(\vec{\theta})}, \\
\left< \rho_1(\vec{\theta},t) \rho_1(\vec{\theta}',t) \right> &=&
\frac{F^{-d}}{1-\gamma d} t^{1-\gamma d}
\frac{\delta(\vec{\theta}-\vec{\theta}')}{J(\vec{\theta})},
\end{eqnarray}
if $\gamma d>1$,
\begin{eqnarray}
\left< \rho_1(\vec{\theta},t) \rho_1(\vec{\theta}',s) \right> &=&
\frac{1}{F^{d}} \mathrm{ln} \left( \min\{ t,s \} \right)
\frac{\delta(\vec{\theta}-\vec{\theta}')}{J(\vec{\theta})}, \\
\left< \rho_1(\vec{\theta},t) \rho_1(\vec{\theta}',t) \right> &=&
\frac{1}{F^{d}} \mathrm{ln}(t)
\frac{\delta(\vec{\theta}-\vec{\theta}')}{J(\vec{\theta})},
\end{eqnarray}
if $\gamma d =1$, and finally
\begin{equation}
\left< \rho_1(\vec{\theta},t) \rho_1(\vec{\theta}',s) \right> =
\frac{F^{-d}}{\gamma d-1}t_0^{1-\gamma d}
\frac{\delta(\vec{\theta}-\vec{\theta}')}{J(\vec{\theta})},
\end{equation}
when $\gamma d>1$. In this last case the correlation vanishes in
the limit $t_0 \to \infty$. Note that the reparametrization
invariance principle is not able to capture dilution effects and
it reproduces pure dilatation dynamics.

In order to introduce dilution in the radial case we may use the
following functional definition which transforms
Eq.~(\ref{rdeposition}) into
\begin{equation}
\label{rddilution}
\partial_t r = F \gamma
t^{\gamma-1} -\frac{\gamma d}{t}r + \frac{1}{r^{d/2}}
\frac{\xi(\vec{\theta},t)}{J(\vec{\theta})^{1/2}},
\end{equation}
whose first moment can be exactly calculated again taking
advantage of the It\^{o} interpretation of the noise term,
yielding
\begin{equation}
\left< r(\vec{\theta},t)
\right>=\frac{F}{d+1}t^\gamma.
\end{equation}
Performing as in the former case the small noise expansion
$r=R+\sqrt{\epsilon}\rho_1$ we find again $R= \left< r \right>$.
The perturbation obeys the equation
\begin{equation}
\partial_t \rho_1 = -\frac{\gamma d}{t}\rho_1+ \frac{(d+1)^{d/2}}{F^{d/2}t^{\gamma
d/2}} \frac{\eta(\vec{\theta},t)}{J(\vec{\theta})^{1/2}},
\end{equation}
and so the perturbation has zero mean and its long time
correlation is given by
\begin{equation}
\left< \rho_1(\vec{\theta},t) \rho_1(\vec{\theta}',s) \right>=
\frac{(d+1)^d}{F^d (\gamma d +1)} \min\{s,t\} \max\{s,t\}^{-\gamma
d} \frac{\delta(\vec{\theta} - \vec{\theta}')}{J(\vec{\theta})},
\end{equation}
a result that holds uniformly in $\gamma$. Note that the structure
of the temporal correlation is different when the effect of
dilution is considered and when it is not for all $\gamma>0$. For
instance, the characteristic length scale corresponding to a given
angular difference is $\mathfrak{l} = \max\{s,t\}^{\gamma}
|\vec{\theta} - \vec{\theta}'|$ when dilution is present, and
$\mathfrak{l} = \min\{s,t\}^{\gamma} |\vec{\theta} -
\vec{\theta}'|$ in the absence of dilution. One already sees in
this example that the lack of dilution causes the appearance of
memory effects on the growth dynamics. The first order correction
in the small noise expansion $\rho_1$ is always a Gaussian
stochastic process; an attempt to go beyond Gaussianity by
deriving the second order correction is reported in appendix
\ref{horder}.

\begin{figure}[h]
\begin{center}
\includegraphics[scale=0.3]{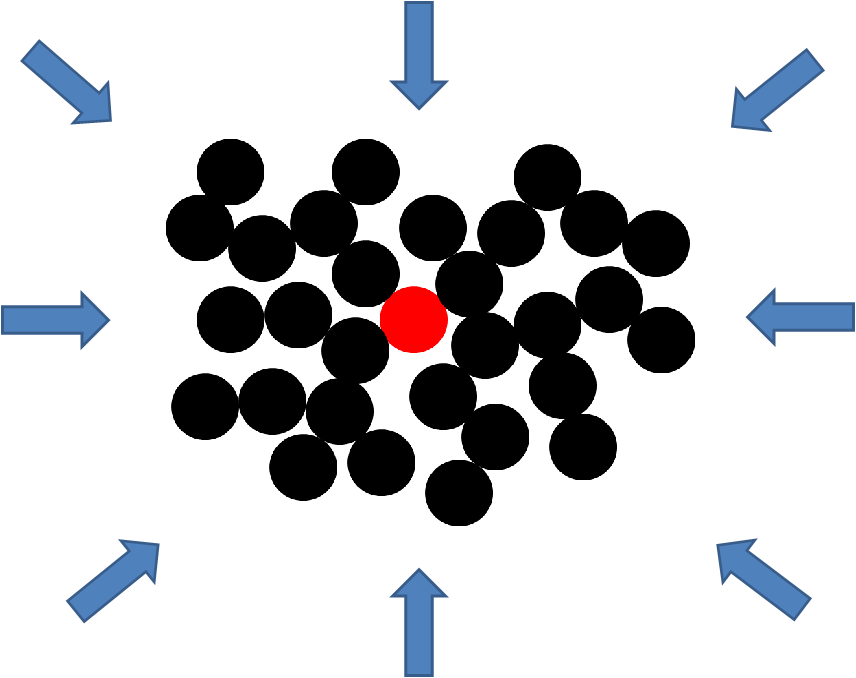}
\caption{Possible realization of the random deposition process as
explained in the text. New particles approach the origin (which is
coincident with the center of the red particle) following a random
direction. Once these particles touch another particle in the
cluster they become permanently attach to that position.}
\label{diagram}
\end{center}
\end{figure}

One may wonder about different possible realizations of the radial
random deposition process we have introduced by means of
stochastic differential equations. It is of particular usefulness
within the realm of statistical physics the design of suitable
discrete models. One possibility would be the proposition sketched
in Fig.~(\ref{diagram}), which focus on a two-dimensional system.
We consider an off-lattice model to get rid of possible
undesirable anisotropy effects which may arise when an underlying
lattice is present. We start placing the center of a spherical
particle at the origin; this is the seed particle colored in red
in Fig.~(\ref{diagram}). The next step would be choosing a random
number uniformly distributed in the interval $[0,2 \pi)$. This
number selects the angle which marks the direction along which a
new particle (identical to the seed one) travels from a long
distance (much longer than the cluster radius) towards the origin.
This particle does not deviate in any sense from this direction
until it touches the seed particle, to which it becomes
permanently attached. The process is now repeated, a new angle is
randomly selected and a new particle follows the corresponding
direction until it touches any other particle in the system; in
this instant it becomes again permanently attached to that
position. Iterating this process we arrive to a growing cluster
like the one depicted in Fig.~(\ref{diagram}). This sort of
cluster, in the limit in which the cluster radius is much longer
than the particles radius, constitutes a possible realization of
the radial random deposition process. The generalization to
arbitrary dimensions is straightforward. In the case of a $d+1$
dimensional cluster the direction along which a new particle
approaches the growing cluster is selected by $d$ random angles,
$d-1$ of them (the polar angles) are uniformly distributed in
$[0,\pi]$ and the $d-$th angle (the azimuth angle) is uniformly
distributed in $[0,2\pi)$. The dynamics of the process is
otherwise identical. We note this discrete model can be understood
as a particular limit of a DLA process in which diffusion is
substituted by a random drift.

\begin{figure}
\centering \subfigure[]{
\includegraphics[width=0.4\textwidth]{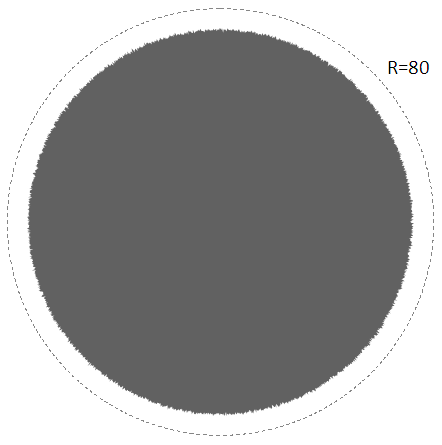}
\label{morpho2}} \subfigure[]{
\includegraphics[width=0.4\textwidth]{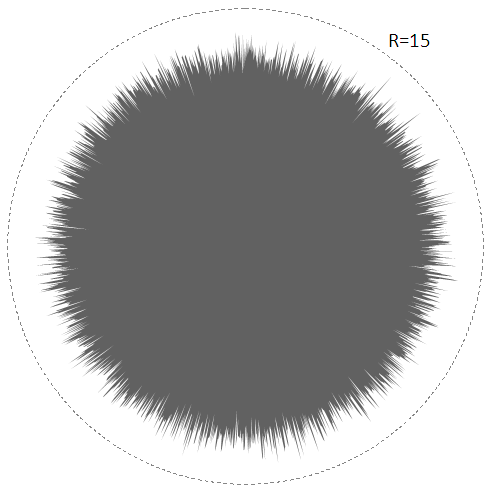}
\label{morpho1}} \subfigure[]{
\includegraphics[width=0.4\textwidth]{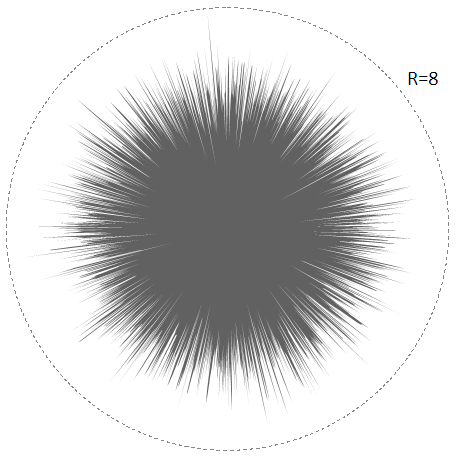}
\label{morpho3}} \subfigure[]{
\includegraphics[width=0.4\textwidth]{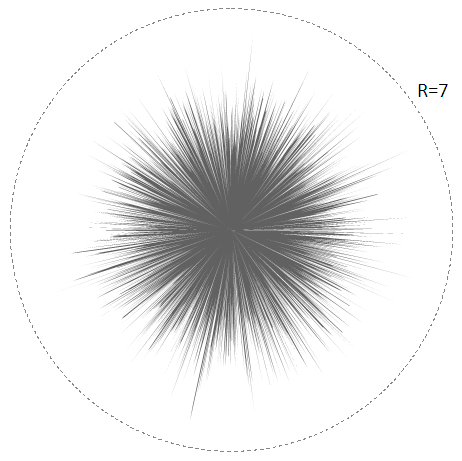}
\label{morpho4} } \caption{Radial random deposition growth
processes plotted at $t=6$. The values of the parameters are
$F=2$, $\epsilon=(200\pi)^{-1}$, and the system has been
discretized in $2^{11}$ spatial points. We have used as initial
condition $r(\theta,1)=2$ (note that we have used as initial time
$t_0=1$). The outer dashed circumferences give an estimate of the
size of the cluster (the corresponding values of the radius are
indicated in each panel). Panel \subref{morpho2} $\gamma=2$. Panel
\subref{morpho1} $\gamma=1$. Panel \subref{morpho3} $\gamma=1/2$.
Panel \subref{morpho4} $\gamma=1/4$.} \label{morphologies}
\end{figure}

Radial random deposition models are important because they act as
statistical attractors of the solutions to different radial growth
equations. Indeed, if the growth index is large enough, then the
long-time large-scale properties of the solutions to different
radial equations approach the corresponding properties of the
solutions to the radial random deposition equations. This has been
characterized as the decorrelation limit in~\cite{escuderojs}. It
is thus interesting to visualize the solutions to these equations.
This is carried out in Fig.~\ref{morphologies}. In this figure we
represent the function $r(\vec{\theta},t)$ from Eq.~(\ref{snoise})
for four different values of $\gamma$ and for $d=1$. The result is
four clearly different morphologies, which become more similar to
the radially symmetric deterministic growth process R(t) for
larger $\gamma$.

\section{Random Deposition and Diffusion}
\label{rdad}

Our next step, in order to approach more complex and realistic
growth processes, is to add diffusion to a random deposition
equation of growth. This sort of equations may be derived using
reparametrization invariance as in~\cite{escudero2}. Following
this reference and the former section, we perform a small noise
expansion and concentrate on the equation for the Gaussian
perturbation. Again, our goal will be deriving for the first time
two-point space-time correlation functions for this type of
processes. In this section we will consider a number of cases
which do not show instabilities, and the study of these will be
postponed to the next one. The equation for the perturbation in
$d=1$ is~\cite{escudero2}
\begin{equation}
\label{perturbation}
\partial_t \rho = \frac{D_\zeta}{(Ft^\gamma)^\zeta} \Lambda_\theta^\zeta \rho + \frac{1}{\sqrt{Ft^\gamma}}\eta(\theta,t),
\end{equation}
where $\Lambda_\theta^\zeta$ is a fractional differential operator
of order $\zeta$, and dilution has not been considered. The
dynamics for $\zeta>d$, which in turn implies that in the linear
case the growth exponent $\beta > 0$ and the interface is
consequently rough, has been already considered
in~\cite{escuderojs}; herein we move to studying the marginal case
$\zeta=d$, which turns out to have interesting properties. The
case $\zeta < d$ is not so interesting as it corresponds to flat
interfaces; an analogous calculation to the corresponding one
in~\cite{escudero2} for $\gamma = 1$ and $\zeta < 1$ shows
\begin{equation}
\left< \rho(\theta,t)\rho(\theta',s) \right> \to 0 \qquad
\mathrm{when} \qquad t,s \to \infty,
\end{equation}
independently of the value of $t_0$.

If $\zeta=\gamma=1$ the correlation reads
\begin{equation}
\left< \rho(\theta,t)\rho(\theta',s) \right>=\frac{1}{4 \pi
D}\mathrm{ln}\left[
\frac{(ts)^{D/F}}{(s/t)^{D/F}+(t/s)^{D/F}-2\cos(\theta-\theta')}
\right].
\end{equation}
The one time correlation adopts the form
\begin{equation}
\left< \rho(\theta,t)\rho(\theta',t) \right>=\frac{1}{4 \pi D}
\mathrm{ln}\left[ \frac{t^{2D/F}}{2-2\cos \left( \theta-\theta'
\right)} \right],
\end{equation}
that reduces to
\begin{equation}
\label{onetime} \left< \rho(\theta,t)\rho(\theta',t) \right>
\approx \frac{1}{2 \pi F} \mathrm{ln}\left( \frac{t}{\left|
\theta-\theta' \right|^{F/D}} \right),
\end{equation}
when we consider local in space dynamics, this is, in the limit
$\theta \approx \theta'$. Note that this result allows us to
define the local dynamic exponent $z_{loc}=F/D \in (0,\infty)$,
which depends continuously on the equation parameters $F$ and $D$,
and is thus nonuniversal, as we noted in~\cite{escudero2}. In
terms of the arc-length variable $\ell-\ell'=t(\theta -\theta')$
we find
\begin{equation}
 \left< \rho(\ell,t)\rho(\ell',t) \right>
\approx \frac{F^{-1}+D^{-1}}{2 \pi} \mathrm{ln}\left(
\frac{t}{\left| \ell-\ell' \right|^{F/(D+F)}} \right),
\end{equation}
where the dynamical exponent in terms of the arc-length variable
$z_\ell =F/(D+F) \in (0,1)$ is again nonuniversal. If we take into
account dilution Eq. (\ref{perturbation}) transforms to
\begin{equation}
\partial_t \rho = \frac{D}{Ft} \Lambda_\theta \rho -\frac{1}{t}\rho + \frac{1}{\sqrt{Ft}}\eta(\theta,t).
\end{equation}
The solution has zero mean and its correlation is given by
\begin{eqnarray}
\left< \rho(\theta,t)\rho(\theta',s) \right> =
\frac{\min\{s,t\}/\max\{s,t\}}{4 \pi F} + \hspace{8.5cm} \\
\frac{(\min\{s,t\}/\max\{s,t\})^{1+D/F}}{2 \pi (F+D)} \Re \left\{
e^{i(\theta-\theta')} {_2F_1}\left[ 1,1+\frac{F}{D};2+\frac{F}{D};
e^{i(\theta-\theta')} \left( \frac{\min\{s,t\}}{\max\{s,t\}}
\right)^{D/F} \right] \right\}, \nonumber
\end{eqnarray}
where $\Re(\cdot)$ denotes the real part and
${_2F_1}(\cdot,\cdot;\cdot;\cdot)$ is Gauss hypergeometric
function~\cite{stegun}. This correlation, for $s=t$ and for small
angular scales $\theta \approx \theta'$, becomes at leading order
\begin{equation}
\left< \rho(\theta,t)\rho(\theta',t) \right> \approx \frac{-1}{2
\pi D} \mathrm{ln} \left(|\theta-\theta'| \right),
\end{equation}
which is time independent, and for the arc-length variable
\begin{equation}
\left< \rho(\ell,t)\rho(\ell',t) \right> \approx \frac{1}{2 \pi D}
\mathrm{ln}\left(\frac{t}{|\ell-\ell'|}\right),
\end{equation}
for which the planar scaling and the universal dynamical exponent
$z=1$ are recovered, see Eq. (C5) in \cite{escudero2}. This is yet
another example, this time of a different nature, of how dilution
is able to restore the Family-Vicsek
scaling~\cite{escuderojs,escuderoar}.

If $\zeta=1$ and $\gamma < 1$ we find the following correlation
function
\begin{eqnarray}
\nonumber \left< \rho(\theta,t)\rho(\theta',s)
\right>=\frac{[\min\{t,s\}]^{1-\gamma}}{2 \pi F
(1-\gamma)}-\frac{1}{4 \pi D}
\mathrm{ln}\left\{1+\exp \left[ -\frac{2D}{F(1-\gamma)} \left|t^{1-\gamma}-s^{1-\gamma} \right| \right] \right. \\
\left. -2 \exp \left[ -\frac{D}{F(1-\gamma)}
\left|t^{1-\gamma}-s^{1-\gamma} \right|
\right]\cos(\theta-\theta') \right\}.
\end{eqnarray}
When $t=s$ we get
\begin{equation}
\left< \rho(\theta,t)\rho(\theta',t) \right>=
\frac{t^{1-\gamma}}{2 \pi F(1-\gamma)}-\frac{1}{4 \pi D} \ln
\left[ 2-2\cos \left( \theta-\theta' \right) \right],
\end{equation}
and considering local spatial dynamics we arrive at
\begin{equation}
\label{onetime2} \left< \rho(\theta,t)\rho(\theta',t) \right>
\approx \frac{t^{1-\gamma}}{2 \pi F(1-\gamma)}-\frac{1}{2 \pi D}
\mathrm{ln} \left( \left| \theta-\theta' \right| \right)=
\frac{1}{2 \pi F (1-\gamma)} \mathrm{ln} \left[
\frac{e^{t^{1-\gamma}}}{|\theta-\theta'|^{F(1-\gamma)/D}} \right],
\end{equation}
expression that does not allow to define a local dynamic exponent,
or alternatively $z_{loc}=0$ due to the exponentially fast
spreading of the correlations. These last three expressions
contain two clearly different terms. The first one is the zeroth
mode component of the correlation, which does not achieve long
time saturation. The second term is the nontrivial stationary part
of the correlation generated along the evolution. As can be seen,
both spatial and temporal correlations are generated.

When the dilution term is taken into account we find the
correlation
\begin{eqnarray}
\nonumber \left< \rho(\theta,t)\rho(\theta',s) \right> =
\frac{\min\{t,s\}[\max \{t,s\}]^{-\gamma}}{2 \pi F (\gamma
+1)}-\frac{1}{4 \pi D}
\mathrm{ln}\left\{1+\exp \left[ -\frac{2D}{F(1-\gamma)} \left|t^{1-\gamma}-s^{1-\gamma} \right| \right] \right. \\
\left. -2 \exp \left[ -\frac{D}{F(1-\gamma)}
\left|t^{1-\gamma}-s^{1-\gamma} \right|
\right]\cos(\theta-\theta') \right\}. \,\,\,\,
\end{eqnarray}
When $t=s$ we get
\begin{equation}
\left< \rho(\theta,t)\rho(\theta',t) \right>=
\frac{t^{1-\gamma}}{2 \pi F(\gamma +1)}-\frac{1}{4 \pi D} \ln
\left[ 2-2\cos \left( \theta-\theta' \right) \right],
\end{equation}
and considering local spatial dynamics we arrive at
\begin{equation}
\left< \rho(\theta,t)\rho(\theta',t) \right> \approx
\frac{t^{1-\gamma}}{2 \pi F(\gamma +1)}-\frac{1}{2 \pi D}
\mathrm{ln} \left( \left| \theta-\theta' \right| \right)=
\frac{1}{2 \pi F (\gamma +1)} \mathrm{ln} \left[
\frac{e^{t^{1-\gamma}}}{|\theta-\theta'|^{F(\gamma+1)/D}} \right],
\end{equation}
and we see that as in the former case, both prefactor and exponent
are modified, but the still exponentially fast propagation of
correlations implies an effective local dynamical exponent
$z_{loc}=0$. Note that for $\gamma > 1$ a radial random deposition
behavior for large spatial scales is recovered.

Now we move onto the two-dimensional setting. As in the
one-dimensional case we focus on the marginal situation
$d=\zeta=2$, which leads us to denominate this sort of equations
as spherical Edwards-Wilkinson (EW) equations, and $0< \gamma \le
1/2$, as greater values of the growth index lead again to
decorrelation. The straightforward generalization of
Eq.~(\ref{perturbation}) is
\begin{equation}
\label{SEW1}
\partial_t \rho = \frac{K}{(Ft^\gamma)^2} \nabla^2 \rho + \frac{1}{Ft^\gamma \sqrt{\sin(\theta)}}\eta(\theta,\phi,t),
\end{equation}
where the noise is a Gaussian random variable of zero mean and
correlation given by
\begin{equation}
\left< \xi(\theta,\phi,t)\xi(\theta',\phi',s)
\right>=\delta(\theta-\theta')\delta(\phi-\phi')\delta(t-s).
\end{equation}
In this case, if $\gamma < 1/2$, the random variable $\rho$ is a
zero mean Gaussian process whose correlation is given by
\begin{eqnarray}
\nonumber
\left< \rho(\theta,\phi,t)\rho(\theta',\phi',s) \right>=\frac{\left[ \min(t,s) \right]^{1-2\gamma}}{4\pi F^2 (1-2\gamma)}+ \\
\label{onetime3} \sum_{l=1}^\infty \sum_{m=-l}^l
\frac{(-1)^m}{2K(l+l^2)}\exp
\left[-\frac{K(l+l^2)}{F^2(1-2\gamma)}\left|t^{1-2\gamma}-
s^{1-2\gamma}\right|\right]Y_{-m}^l(\theta,\phi)Y_m^l(\theta',\phi'),
\end{eqnarray}
where the expansion has been performed on the spherical harmonics
basis $Y_{m}^l(\theta,\phi)$. If $\gamma = 1/2$ then $\rho$
becomes a zero mean Gaussian random variable with the new
correlation
\begin{eqnarray}
\nonumber
\left< \rho(\theta,\phi,t)\rho(\theta',\phi',s) \right>=\frac{\mathrm{ln}\left[ \min(t,s) \right]}{4\pi F^2}+ \\
\label{corrmar} \sum_{l=1}^\infty \sum_{m=-l}^l
\frac{(-1)^m}{2K(l+l^2)}\left[
\frac{\min(s,t)}{\max(s,t)}\right]^{K(l+l^2)/F^2}
Y_{-m}^l(\theta,\phi)Y_m^l(\theta',\phi').
\end{eqnarray}
It is clear that these correlations are again composed of two
different terms, the first one associated with the $l=0$ mode
never saturates, and the second one associated with the rest of
modes $l>0$, which saturates and is responsible of a non-trivial
spatial structure.

Taking into account dilution we find for $\gamma < 1/2$ the
correlation
\begin{eqnarray}
\nonumber \left< \rho(\theta,\phi,t)\rho(\theta',\phi',s)
\right>=\frac{ \min(t,s) \left[ \max(t,s) \right]^{-2\gamma}}{4\pi
F^2 (2\gamma +1)}+ \\
\sum_{l=1}^\infty \sum_{m=-l}^l \frac{(-1)^m}{2K(l+l^2)}\exp
\left[-\frac{K(l+l^2)}{F^2(1-2\gamma)}\left|t^{1-2\gamma}-
s^{1-2\gamma}\right|\right]Y_{-m}^l(\theta,\phi)Y_m^l(\theta',\phi'),
\end{eqnarray}
and for $\gamma=1/2$
\begin{equation}
\left< \rho(\theta,\phi,t)\rho(\theta',\phi',s) \right>=
\sum_{l=0}^\infty \sum_{m=-l}^l
\frac{(-1)^m}{2F^2+2K(l^2+l)}\left[ \frac{\min(s,t)}{\max(s,t)}
\right]^{1+K(l^2+l)/F^2}
Y_{-m}^l(\theta,\phi)Y_m^l(\theta',\phi').
\end{equation}
In the two dimensional situation we see that dilution also has a
measurable effect, which is more pronounced in the critical
$\gamma=1/2$ case. For this value all the modes in the correlation
saturate and contribute to create a stationary spatial structure,
as in the one-dimensional setting. It is difficult to establish
more comparisons among both dimensionalities, as the infinite sums
that were explicit in $d=1$ become much more involved in $d=2$,
due to the double series containing the spherical harmonics. We
nevertheless conjecture that the modification of the scaling
properties due to effect of dilution in two dimensions is similar
to the one explicitly observed in one dimension.

As in the last section we may wonder what would be the simplest
realizations of these processes in the form of discrete systems.
Due to the generality of this question we are not in position to
yield a complete answer to it in this moment. We can however point
to a direction which seems promising. Several discrete models are
known to be in the theoretical university classes described by
stochastic growth equations in the classical situation of planar
static domains~\cite{barabasi}. These models can be cast on a
growing, still planar, domain following the technique employed
in~\cite{galeano}. This would a first step in the search for
suitable discrete radial models. We note the agreement of the
computational results in~\cite{galeano} with our theoretical
predictions.

\section{Instabilities}
\label{instabilities}

In this section we will analyze the effect that instabilities have
on the two-point space-time correlation functions calculated for
the last section processes.

A spherical EW equation derived from the geometric principle of
surface minimization was introduced in~\cite{escudero2}. The
corresponding equation for the radius $r(\theta,\phi,t)$ reads
\begin{equation}
\label{SEW2}
\partial_t r = K \left[ \frac{\partial_\theta r}{r^2 \tan(\theta)}+
\frac{\partial_\theta^2 r}{r^2}+\frac{\partial_\phi^2 r}{r^2
\sin^2(\theta)}-\frac{2}{r}\right] + F\gamma t^{\gamma-1}
+\frac{1}{r\sqrt{\sin(\theta)}}\xi(\theta,\phi,t).
\end{equation}
Performing the small noise expansion
$r(\theta,\phi,t)=Ft+\rho(\theta,\phi,t)$ we find a linear
equation which differs from Eq. (\ref{SEW1}) in that it has a
destabilizing term coming from the fourth term in the drift of Eq.
(\ref{SEW2}), see \cite{escudero2}. In this reference one can see
that in the absence of dilution the $l=0$ mode is unstable and the
$l=1$ modes are marginal while the rest of modes is stable. The
effect of this sort of geometrically originated instability on the
mean value of the stochastic perturbation and alternative
geometric variational approaches that avoid it can be seen
in~\cite{escudero2}, herein we will concentrate on its effect on
correlations. Its effect on mean values can be easily deduced from
them.

In the long time limit and provided $\gamma < 1/2$, the
perturbation is a Gaussian process whose correlation is given by
\begin{eqnarray}
\nonumber
\left< \rho(\theta,\phi,t)\rho(\theta',\phi',s) \right>=\frac{1}{16\pi K}\exp \left[ \frac{2K(t^{1-2\gamma}+s^{1-2\gamma})}{F^2(1-2\gamma)}\right]+ \\
\nonumber
\frac{3\left[ \min(t,s) \right]^{1-2\gamma}}{4\pi F^2 (1- 2\gamma)}\left[\cos(\theta)\cos(\theta')+\cos(\phi-\phi') \sin(\theta)\sin(\theta')\right]+ \\
\sum_{l=2}^\infty \sum_{m=-l}^l
\frac{(-1)^m}{2K(l^2+l-2)}\exp\left[-\frac{K(l^2+l-2)}{F^2(1-2\gamma)}
\left|t^{1-2\gamma}-s^{1-2\gamma}\right|\right]Y_{-m}^l(\theta,\phi)Y_m^l(\theta',\phi').
\label{unscorr1}
\end{eqnarray}
If $\gamma = 1/2$ the correlation shifts to
\begin{eqnarray}
\nonumber
\left< \rho(\theta,\phi,t)\rho(\theta',\phi',s) \right>=\frac{(st/t_0^2)^{2K/F^2}}{16\pi K}+ \\
\nonumber
\frac{3\mathrm{ln}\left[ \min(s,t) \right]}{4\pi F^2}\left[\cos(\theta)\cos(\theta')+\cos(\phi-\phi') \sin(\theta)\sin(\theta')\right]+ \\
\sum_{l=2}^\infty \sum_{m=-l}^l \frac{(-1)^m}{2K(l^2+l-2)}\left[
\frac{\min(s,t)}{\max(s,t)} \right]^{K(l^2+l-2)/F^2}
Y_{-m}^l(\theta,\phi)Y_m^l(\theta',\phi'). \label{unscorr2}
\end{eqnarray}
In these cases the modes characterized by $l=0$ and $l=1$ do not
saturate, and the rest of the modes $l>1$ saturate and create a
non-trivial spatial structure. When $\gamma < 1/2$ the $l=1$ modes
grow in time as a power law with the exponent $1-2\gamma$, while
the $l=0$ mode grows exponentially fast. When $\gamma=1/2$ the
$l=1$ modes grow logarithmically and the $l=0$ mode grows as a
power law with the non-universal exponent $4K/F^2$.

When we consider the effect of dilution, and for $\gamma < 1/2$,
we find the correlation
\begin{eqnarray}
\nonumber
\left< \rho(\theta,\phi,t)\rho(\theta',\phi',s) \right>=\frac{1}{16\pi K}\exp \left[ \frac{2K(t^{1-2\gamma}+s^{1-2\gamma})}{F^2(1-2\gamma)}\right]+ \\
\nonumber
\frac{3 \min(t,s) \left[ \max(t,s) \right]^{-2\gamma}}{4\pi F^2 (2\gamma +1)}\left[\cos(\theta)\cos(\theta')+\cos(\phi-\phi') \sin(\theta)\sin(\theta')\right]+ \\
\sum_{l=2}^\infty \sum_{m=-l}^l
\frac{(-1)^m}{2K(l^2+l-2)}\exp\left[-\frac{K(l^2+l-2)}{F^2(1-2\gamma)}
\left|t^{1-2\gamma}-s^{1-2\gamma}\right|\right]Y_{-m}^l(\theta,\phi)Y_m^l(\theta',\phi').
\end{eqnarray}
For $\gamma=1/2$ the correlation reads
\begin{eqnarray}
\nonumber
\left< \rho(\theta,\phi,t)\rho(\theta',\phi',s) \right>=\frac{1}{4 \pi} \left< \rho_0^0(t) \rho_0^0(s) \right> + \\
\sum_{l=1}^\infty \sum_{m=-l}^l
\frac{(-1)^m}{2F^2+2K(l^2+l-2)}\left[ \frac{\min(s,t)}{\max(s,t)}
\right]^{1+K(l^2+l-2)/F^2}
Y_{-m}^l(\theta,\phi)Y_m^l(\theta',\phi'),
\end{eqnarray}
where
\begin{equation}
\left< \rho_0^0(t) \rho_0^0(s) \right> = \left\{
\begin{array}{lll}
(2F^2-4K)^{-1} \left( \min\{s,t\}/\max\{s,t\} \right)^{1-2K/F^2} & \mbox{\qquad if \qquad $F^2>2K$}, \\
\mathrm{ln}(\min\{s,t\})/F^2 & \mbox{\qquad if \qquad
$F^2=2K$}, \\
(4K-2F^2)^{-1} \left( t s/t_0^2 \right)^{2K/F^2-1} & \mbox{\qquad
if \qquad $F^2<2K$},
\end{array} \right.
\end{equation}
where $t_0$ is the absolute origin of time.

Contrary to what happens in the stable case, Eq.~(\ref{SEW1}), in
the unstable case with no dilution, Eq.~(\ref{SEW2}), the $l=0$
mode is unstable, showing an exponential growth, and the $l=1$
modes shows an algebraic increase with the universal exponent
$1-2\gamma$, provided $\gamma<1/2$; the rest of modes is stable.
The marginal value of the growth index $\gamma = 1/2$ translates
into a power law increase of the $l=0$ mode with a non-universal
exponent, while the $l=1$ modes grow logarithmically; the rest of
modes is again stable. It is clear that dilution has a stabilizing
effect. Indeed, for $\gamma<1/2$ the $l=0$ mode is unchanged, but
the $l=1$ modes, which still grow in time, experience a lost of
memory effects. In the critical $\gamma=1/2$ situation the
dilution effects are stronger. The $l=1$ modes, which formerly
grew logarithmically, now become stable; the $l=0$ mode, which
formerly showed an algebraic growth, now shows (non-universal)
algebraic or logarithmic grow, or even saturation, depending on
the relation among the values of the parameters of the spherical
EW equation. In any case, even that of algebraic growth, this
growth is always slower than in the no dilution situation. Stable
modes saturate contributing to create a non-trivial spatial
structure in the whole range $\gamma \le 1/2$.

In summary, the effect of dilution is weakly stabilizing in the
subcritical case, while stronger and more identifiable in
criticality. Of course, the supercritical situation is
characterized by an effective random deposition behavior in the
large spatial scale.

\section{Intrinsically spherical growth and rapid roughening}
\label{rapidrough}

In this section we will show for the first time how rapid
roughening naturally appears in the radial growth setting. To this
end it is necessary to clarify the role of the diffusivity index
$\zeta$. We have defined it as the order of the fractional
differential operator taking mass diffusion into account, and so
far we have referred to it as the key element triggering
decorrelation. This has been an abuse of language because we have
assumed that the negative power of the radius (or its mean field
analog $Ft^\gamma$ -- what really matters is the resulting power
of the temporal variable) preceding this differential operator was
exactly $\zeta$. This would not be the case if the diffusion
constant were time or radius dependent, but also in some other
cases as the Intrinsically Spherical (IS) equation derived from
geometric variational principles in~\cite{escudero2}. This
equation was obtained as a gradient flow pursuing the minimization
of the interface mean curvature, and then linearizing with respect
to the different derivatives of the radius as given by the small
gradient assumption~\cite{escudero2}. It is termed ``intrinsically
spherical'' because it has no planar counterpart, as the
nonlinearity becomes fundamental in any attempt to derive such a
gradient flow in the Cartesian framework~\cite{escudero3,elka}.
Note the similarity of this with other equations classical in this
context, as the EW equation is a gradient flow which minimizes the
surface area and the Mullins-Herring equation minimizes the
interface square mean curvature~\cite{escudero2}; the IS equation,
as mentioned, minimizes the interface mean curvature. It
reads~\cite{escudero2}
\begin{equation}
\partial_t r= K \left[ \frac{\partial_\theta^2 r}{r^3}+\frac{\partial_\phi^2 r}{r^3 \sin^2(\theta)}
+\frac{\partial_\theta r}{r^3 \tan(\theta)}-\frac{1}{r^2}\right] +
F\gamma t^{\gamma-1}
+\frac{1}{r\sqrt{\sin(\theta)}}\xi(\theta,\phi,t),
\end{equation}
and so $\zeta=2$ in this case, but however one finds a factor
$r^{-3}$ in front of the diffusive differential operator, instead
of the $r^{-2}$ factor characteristic of the EW equation. This
difference will have a number of measurable consequences, as we
will show in the following. The equation for the stochastic
perturbation reads in this case
\begin{equation}
\frac{d \rho^l_m}{dt}=\frac{K}{F^3 t^{3 \gamma}}[2-l(l+1)]\rho^l_m
-\frac{2 \gamma}{t}\rho^l_m + \frac{1}{Ft^\gamma}\eta^l_m(t),
\end{equation}
which reveals that the critical value of the growth index
$\gamma=1/3$; a faster growth leads to decorrelation. This is the
first but not the unique difference with respect to the EW
equation. To find out more we will first put things in a broader
context.

A more general equation for radial growth, after introducing
dilution, would be
\begin{equation}
\label{damping}
\partial_t r= -\frac{K}{r^\delta} |\nabla|^\zeta r -\frac{\gamma
d}{t}r + F\gamma t^{\gamma-1} +\frac{\sqrt{\epsilon}}{\sqrt{r^d
J(\vec{\theta})}}\eta(\vec{\theta},t),
\end{equation}
which defines the damping index $\delta$, differing from the
diffusivity index $\zeta$ in general; note that
Eq.~(\ref{damping}) has left aside the instability properties of
the IS equation, which are analogous to those of the EW equation,
and would add nothing to last section discussion. For simplicity
we will focus on values of the damping index fulfilling $\delta
\ge \zeta$. This equation can be treated perturbatively for small
$\epsilon$ following the previous sections procedure and by
introducing the hyperspherical harmonics
$Y^{\vec{m}}_l(\vec{\theta})$, which obey the eigenvalue
equation~\cite{wen}
\begin{equation}
\nabla^2 Y^{\vec{m}}_l(\vec{\theta})=
-l(l+d-1)Y^{\vec{m}}_l(\vec{\theta}),
\end{equation}
where the vector $\vec{m}$ represents the set of $(d-1)$ indices.
The fractional operator acts on the hyperspherical harmonics in
the following fashion
\begin{equation}
|\nabla|^\zeta Y^{\vec{m}}_l(\vec{\theta})=
[l(l+d-1)]^{\zeta/2}Y^{\vec{m}}_l(\vec{\theta}).
\end{equation}
The hyperspherical noise is Gaussian, has zero mean and its
correlation is given by
\begin{equation}
\left< \eta(\vec{\theta},t) \eta(\vec{\theta}',t') \right>=
\delta(\vec{\theta}-\vec{\theta}') \delta(t-t').
\end{equation}
It can be expanded in terms of hyperspherical harmonics
\begin{equation}
\frac{\eta(\vec{\theta},t)}{\sqrt{J(\vec{\theta})}}=\sum_{l,\vec{m}}
\eta_l^{\vec{m}}(t) Y^{\vec{m}}_l(\vec{\theta}),
\end{equation}
and the amplitudes are given by
\begin{equation}
\eta_l^{\vec{m}}(t)=\int \eta(\vec{\theta},t)
\bar{Y}^{\vec{m}}_l(\vec{\theta}) \sqrt{J(\vec{\theta})} \,
d\vec{\theta},
\end{equation}
and so they are zero mean Gaussian noises whose correlation is
given by
\begin{equation}
\left< \eta_l^{\vec{m}}(t) \bar{\eta}_{l'}^{\vec{m}'}(t')
\right>=
\delta(t-t') \delta_{l,l'} \delta_{\vec{m},\vec{m}'},
\end{equation}
where the overbar denotes complex conjugation. Note that the
amplitudes are in general complex valued. They obey the linear
equation
\begin{equation}
\frac{d \rho_l^{\vec{m}}}{dt}=-\frac{K}{F^\delta t^{\delta
\gamma}}[l(l+d-1)]^{\zeta/2}\rho_l^{\vec{m}} -\frac{\gamma
d}{t}\rho_l^{\vec{m}} +\frac{1}{F^{d/2}t^{\gamma d/2}}
\eta_l^{\vec{m}}(t).
\end{equation}
From this equation it is clear that the critical value of the
growth index is $\gamma=1/\delta$, and a faster growth leads to
decorrelation.

It is convenient to move to a growing hypercubic geometry as
in~\cite{escuderojs} in order to calculate different quantities
\begin{equation}
\label{hcubic}
\partial_t h=-D \left( \frac{t_0}{t} \right)^{\delta \gamma} |\nabla|^\zeta h
-\frac{d\gamma}{t}h +\gamma F t^{\gamma-1}+
\left(\frac{t_0}{t}\right)^{d\gamma/2}\xi(x,t),
\end{equation}
since this change simplifies calculations without modifying the
leading results. Our goal is finding the growth and
auto-correlation exponents, as this last one is a good quantity to
measure decorrelation \cite{escuderojs}. In order to calculate the
temporal correlations we need to consider the short time limit,
where the growth exponent $\beta$ becomes apparent. The propagator
of Eq.(\ref{hcubic}) is
\begin{equation}
G_n(t)= \left(\frac{t}{t_0}\right)^{-d\gamma} \exp \left[
-\frac{n^\zeta \pi^\zeta D}{L_0^\zeta} \frac{t_0^{\gamma \delta}
t^{1-\gamma \delta}-t_0}{1-\gamma \delta} \right],
\end{equation}
that yields the following complete solution when the initial
condition vanishes:
\begin{equation}
h_n(t)=G_n(t) \int_{t_0}^{t} G_n^{-1}(\tau)
\left(\frac{t_0}{\tau}\right)^{d \gamma/2} \xi_n(\tau) d\tau.
\end{equation}
The one point two times correlation function then reads
\begin{equation}
\label{corrf} \langle h_n(t)h_n(t') \rangle \sim G_n(t) G_n(t')
\int_{t_0}^{\min(t,t')} G_n^{-2}(\tau) \left( \frac{t_0}{\tau}
\right)^{d \gamma} d\tau,
\end{equation}
and after inverting Fourier we arrive at the real space expression
\begin{equation}
\label{realseries} \langle h(x,t)h(x,t') \rangle =
\sum_{n=0}^\infty \langle h_n(t) h_n(t') \rangle \cos^2 \left(
\frac{n \pi x}{L_0} \right),
\end{equation}
where we have assumed no flux boundary conditions as in
\cite{escuderojs}, although the values of both the growth and
auto-correlation exponents do not depend on the choice of boundary
conditions. The propagator $G_n(t)$ suggests the scaling variable
$v_n \sim nt^{(1-\gamma \delta)/\zeta}$ in Fourier space, that
corresponds to the real space scaling variable $u \sim
xt^{(-1+\gamma \delta)/\zeta}$, as can be read directly from Eq.
(\ref{realseries}). This suggests the definition of the effective
dynamical exponent $z_{\mathrm{eff}}= \zeta/(1-\gamma \delta)$. If
we express the correlation Eq. (\ref{corrf}) for $t=t'$ in terms
of the scaling variable $v_n$ (and we refer to it as $C(v_n)$
multiplied by a suitable power of $t$) and we introduce the
``differential'' $1 \equiv \Delta n \sim t^{(-1+\gamma
\delta)/\zeta} \Delta v$, we can cast the last expression in the
integral form
\begin{equation}
\label{avezero}
\langle h(x,t)^2 \rangle -\langle h(x,t) \rangle^2
= t^{1-d/\zeta+\gamma d(\delta/\zeta-1)} \int_{v_1}^\infty C(v_n)
\cos^2\left( \frac{v_n \pi u}{L_0} \right) dv_n,
\end{equation}
where the series converges as a Riemann sum to the above integral
when
\begin{equation}
D t \ll (L_0^\zeta+D t_0){t^{\delta \gamma} \over t_0^{\delta
\gamma}},
\end{equation}
or equivalently $t \ll t_c \sim L_0^{z_{\mathrm{eff}}}$, for $t_c$
being the time it takes the correlations reaching the substrate
boundaries, assuming that the substrate initial size is very
large. If $\gamma <1/\delta$, the whole substrate becomes
correlated, yielding a finite $t_c$; for $\gamma > 1/\delta$ the
convergence of the Riemann sum to the integral is assured for all
times, corresponding to the physical fact that the substrate never
becomes correlated. In front of the integral we find a power of
the temporal variable compatible with the growth exponent
\begin{equation}
\label{beta}
\beta= \frac{1}{2}- \frac{d}{2\zeta}+ \frac{\gamma
d}{2}\left( \frac{\delta}{\zeta}-1 \right),
\end{equation}
and the integral can be shown to be absolutely convergent as the
integrand decays faster than exponentially for large values of the
scaling variable $v_n$.

We are now in position to calculate the temporal auto-correlation
\begin{equation}
\label{temporalc} A(t,t') \equiv \frac{\langle
h(x,t)h(x,t')\rangle_0}{\langle h(x,t)^2 \rangle_0^{1/2} \langle
h(x,t')^2 \rangle_0^{1/2}} \sim
\left(\frac{\min\{t,t'\}}{\max\{t,t'\}}\right)^\lambda,
\end{equation}
where $\lambda$ is the auto-correlation exponent and $\langle
\cdot \rangle_0$ denotes the average with the zeroth mode
contribution suppressed, as in (\ref{avezero}). The remaining
ingredient is the correlation $\langle h(x,t)h(x,t')\rangle_0$.
Going back to Eq.(\ref{realseries}) we see that the Fourier space
scaling variable now reads
\begin{equation}
v_n=\left[\frac{t^{1-\gamma \delta} +(t')^{1-\gamma \delta} - 2
\tau^{1-\gamma \delta}}{1-\gamma \delta} \right]^{1/\zeta}n.
\end{equation}
If $\gamma < 1/\delta$ the term $\max\{t,t'\}^{1-\gamma \delta}$
is dominant and the factor in front of the convergent Riemann sum
reads
\begin{equation}
\max\{t,t'\}^{(\delta/\zeta -1)\gamma d- d/\zeta} \min\{t,t'\},
\end{equation}
after the time integration has been performed and in the limit
$\max\{t,t'\} \gg \min\{t,t'\}$. In this same limit, but when
$\gamma > 1/\delta$, the term $\min\{t,t'\}^{1-\gamma \delta}$
becomes dominant and the prefactor reads
\begin{equation}
\max\{ t,t' \}^{-d\gamma} \min\{t,t'\}^{1- d/\zeta+d\gamma
\delta/\zeta}.
\end{equation}
The resulting temporal correlation adopts the form indicated in
the right hand side of (\ref{temporalc}), where
\begin{equation}
\lambda = \left\{ \begin{array}{ll} \beta + d/\zeta +\gamma
d(1-\delta/\zeta) &
\mbox{\qquad if \qquad $\gamma < 1/\delta$}, \\
\beta +\gamma d & \mbox{\qquad if \qquad $\gamma
> 1/\delta$}, \end{array} \right.
\end{equation}
or alternatively
\begin{equation}
\label{lambda}
\lambda= \beta + {d \over z_\lambda},
\end{equation}
where the $\lambda-$dynamical exponent is defined as
\begin{equation}
z_\lambda = \left\{ \begin{array}{ll}
\frac{\zeta}{1+\gamma(\zeta-\delta)} & \mbox{\qquad if \qquad
$\gamma < 1/\delta$},
\\
1/\gamma & \mbox{\qquad if \qquad $\gamma
> 1/\delta$}. \end{array} \right.
\end{equation}
If we disregarded the effect of dilution we would find again Eq.
(\ref{lambda}), but this time
\begin{equation}
z_\lambda = \left\{ \begin{array}{ll} \frac{\zeta}{1-\gamma
\delta}=z_{\mathrm{eff}} & \mbox{\qquad if \qquad $\gamma <
1/\delta$},
\\
\infty & \mbox{\qquad if \qquad $\gamma
> 1/\delta$}. \end{array} \right.
\end{equation}

To further clarify the dynamics we now calculate the scaling form
that the two points correlation function adopts for short spatial
scales $|x-x'| \ll t^{(1-\delta \gamma)/\zeta}$ in the
decorrelated regime. As dilution does not act on such a
microscopic scale, the following results are independent of
whether we contemplate dilution or not. In this case one has
\begin{equation}
\langle h(x,t)h(x',t) \rangle = \sum_{n_1, \cdots, n_d} \langle
h_n^2(t) \rangle \cos \left( \frac{n_1 \pi x_1}{L_0} \right)\cos
\left( \frac{n_1 \pi x_1'}{L_0} \right) \cdots \cos \left(
\frac{n_d \pi x_d}{L_0} \right)\cos \left( \frac{n_d \pi
x_d'}{L_0} \right),
\end{equation}
where $x=(x_1,\cdots,x_d)$ and $n=(n_1,\cdots,n_d)$, and we assume
the rough interface inequality $\zeta > d$ in order to assure the
absolute convergence of this expression. By introducing the
scaling variables $v_i=n_i t^{(1-\delta \gamma)/\zeta}$ and
$u_i=x_i t^{(\gamma \delta-1)/\zeta}$ for $i=1,\cdots,d$ and
assuming statistical isotropy and homogeneity of the scaling form
we find
\begin{equation}
\langle h(x,t)h(x',t) \rangle - \langle h(x,t) \rangle^2  =
|x-x'|^{\zeta-d} t^{\gamma(\delta-d)}\mathcal{F}\left[ |x-x'|
t^{(\delta \gamma -1)/\zeta} \right],
\end{equation}
or in Lagrangian coordinates $|y-y'|=|x-x'|t^\gamma$
\begin{equation}
\langle h(y,t)h(y',t) \rangle - \langle h(y,t) \rangle^2  =
|y-y'|^{\zeta-d} t^{\gamma(\delta-\zeta)}\mathcal{F}\left[
\frac{|y-y'|}{t^{\{1+ \gamma(\zeta-\delta)\}/\zeta}} \right].
\end{equation}
We see that this form is statistically self-affine with respect to
the re-scaling $y \to b y$, $t \to b^z t$, and $h \to b^\alpha h$,
where the critical exponents are
\begin{equation}
\label{critexp}
\alpha=\frac{\zeta-d}{2}+\frac{\zeta}{1+\gamma(\zeta-\delta)}\frac{(\delta-\zeta)\gamma}{2},
\qquad z=\frac{\zeta}{1+\gamma(\zeta-\delta)}.
\end{equation}
Note that the scaling relation $\alpha=\beta z$ holds, where the
growth exponent $\beta$ was calculated in Eq. (\ref{beta}). The
macroscopic decorrelation, which is observed for length scales of
the order of the system size $|x-x'| \approx L_0$, is controlled
by the effective dynamical exponent $z_{\mathrm{eff}}$. When
$\delta>\zeta$ decorrelation might happen at microscopic length
scales $|x-x'| \ll t^{(1-\delta \gamma)/\zeta}$ as well.
Microscopic decorrelation happens in the limit $\delta \to \zeta +
1/\gamma$. For $\delta < \zeta + 1/\gamma$ the interface is
microscopically correlated and the critical exponents take on
their finite values given in Eq. (\ref{critexp}). For $\delta \ge
\zeta + 1/\gamma$ the interface is microscopically uncorrelated
and the critical exponents diverge $\alpha=z=\infty$, while the
growth exponent is still finite and given by Eq. (\ref{beta}) (so
one could say the scaling relation $\alpha= \beta z$ still holds
in some sense in the microscopic uncorrelated limit). With respect
to the growth exponent we can say that $\beta <1/2$ when $\delta <
\gamma^{-1}+\zeta$, $\beta \to 1/2$ when $\delta \to
\gamma^{-1}+\zeta$, and $\beta > 1/2$ when $\delta >
\gamma^{-1}+\zeta$, so rapid roughening is a consequence of
microscopic decorrelation. And now, by applying the developed
theory to the IS equation, for which $d=2$, $\zeta=2$, $\delta=3$
and assuming as in~\cite{escudero2} that $\gamma=1$, we find that
it is exactly positioned at the threshold of microscopic
decorrelation, this is, its critical exponents are
$\alpha=z=\infty$ and $\beta=1/2$.

Note that the effective dynamical exponent
$z_{\mathrm{eff}}=\zeta/(1-\gamma \delta)$ states the speed at
which both correlation and decorrelation occur. The transition
from correlation to decorrelation is triggered by the comparison
among the indexes $\gamma$ and $\delta$. The derivation order
$\zeta$ controls the speed at which both processes happen: a
larger $\zeta$ implies slower correlation/decorrelation processes.
Note also that rapid roughening might appear in exactly the same
way in planar processes, just by allowing field or time dependence
on the diffusion constant. This is actually the case in some
planar situations~\cite{ehg}, and we have also shown that it
appears naturally in the radial case, where such a dependence is a
straightforward consequence of the lost of translation invariance,
due to the existence of an absolute origin of space, characterized
by a zero radius (and which in turn implies the existence of an
absolute origin of time in the small noise approximation, as we
have already seen). Such a naturalness can be seen in the
derivation of the IS equation in~\cite{escudero2}, where it was
found as a consequence of a simple variational principle.

\section{Scale dependent fractality}
\label{polyfractal}

We devote this section to show, for the first time, that rapidly
growing radial interfaces develop {\it scale dependent
fractality}. This expression denotes a behavior characterized by a
scale dependent fractal dimension taking place in a finite system
and for long times. It is different from the concept of
multifractality, which in this topic is usually associated to a
nonlinear relation among the exponents characterizing the higher
order height difference correlations~\cite{barabasi}.

In the classical case of static planar interfaces the fractal
dimension is computed from the height difference correlation
function
\begin{equation}
\left< [h(x,t)-h(x',t)]^2 \right>^{1/2} \sim |x-x'|^H,
\end{equation}
in the long time limit, i. e. after saturation have been achieved,
where the Hurst exponent $H=(\zeta-d)/2$ for linear growth
equations and the right hand side is time independent. The
interface fractional dimension is calculated using the box
counting method and is given by $d_f=1+d-H$. The general linear
equation for stochastic growth on a growing domain was found in
the last section to be
\begin{equation}
\label{hcubic2}
\partial_t h=-D \left( \frac{t_0}{t} \right)^{\delta \gamma} |\nabla|^\zeta h
-\frac{d\gamma}{t}h +\gamma F t^{\gamma-1}+
\left(\frac{t_0}{t}\right)^{d\gamma/2}\xi(x,t),
\end{equation}
for which we will assume $\zeta \le \delta < \zeta + \gamma^{-1}$.
Its Fourier transformed version, for $n \ge 1$, is
\begin{equation}
\label{hcubic2f}
\frac{d h_n}{dt}=-D \left( \frac{t_0}{t}
\right)^{\delta \gamma} \frac{\pi^\zeta |n|^\zeta}{L_0^\zeta} h_n
-\frac{d\gamma}{t}h_n +
\left(\frac{t_0}{t}\right)^{d\gamma/2}\xi_n(t).
\end{equation}
For slow growth $\gamma < 1/\delta$ diffusion dominates over
dilution and one finds an expression compatible with that of the
planar case
\begin{equation}
\left< [h(x,t)-h(x',t)]^2 \right>^{1/2} \sim
t^{\gamma(\delta-d)/2} |x-x'|^{(\zeta-d)/2},
\end{equation}
and so the Hurst exponent and interface fractal dimension are the
same as in the planar case for fixed time. In the case of fast
growth $\gamma>1/\delta$, for small spatial scales $|x-x'| \ll
t^{(1-\delta \gamma)/\zeta}$ we recover again this result, while
for large spatial scales $|x-x'| \gg t^{(1-\delta \gamma)/\zeta}$
we find
\begin{equation}
\left< [h(x,t)-h(x',t)]^2 \right>^{1/2} \sim t^{\beta},
\end{equation}
and so, for fixed time, $H=0$ and $d_f=d+1$. This means that the
interface becomes highly irregular and so dense that it fills the
$(d+1)-$dimensional space. This way decorrelation marks the onset
of scale dependent fractality, as specified by a scale dependent
Hurst exponent, whose asymptotic values are
\begin{equation}
H(|x-x'|,t) = \left\{ \begin{array}{ll} (\zeta-d)/2 & \mbox{\qquad
if \qquad $|x-x'| \ll t^{(1-\delta \gamma)/\zeta}$},
\\
0 & \mbox{\qquad if \qquad $|x-x'| \gg t^{(1-\delta
\gamma)/\zeta}$}, \end{array} \right.
\end{equation}
and the corresponding asymptotic values of the scale dependent
fractal dimension
\begin{equation}
d_f(|x-x'|,t) = \left\{ \begin{array}{ll} 1+(3d-\zeta)/2 &
\mbox{\qquad if \qquad $|x-x'| \ll t^{(1-\delta \gamma)/\zeta}$},
\\
d+1 & \mbox{\qquad if \qquad $|x-x'| \gg t^{(1-\delta
\gamma)/\zeta}$}. \end{array} \right.
\end{equation}
Note that these results imply dynamic scale dependent fractality
as the scale separating the two regimes depends on time $|x-x'|
\sim t^{(1-\delta \gamma)/\zeta}$; also, the rough interface
inequality $\zeta>d$ implies the strict inequality $1+(3d-\zeta)/2
< d+1$. This asymptotic behavior strongly suggests the
self-similar form of both Hurst exponent and fractal dimension
\begin{equation}
H=H \left( \frac{|x-x'|}{t^{(1-\delta \gamma)/\zeta}} \right),
\qquad \mathrm{and} \qquad d_f=d_f \left(
\frac{|x-x'|}{t^{(1-\delta \gamma)/\zeta}} \right).
\end{equation}
According to this the fractal dimension would be a dynamic fractal
itself, invariant to the transformation $x \to b \, x$, $t \to
b^{z_f}t$, and $d_f \to b^{\alpha_f} d_f$, for
$z_f=\zeta/(1-\delta \gamma)=z_{\mathrm{eff}}$ and $\alpha_f=0$.
Note that all these results concerning scale dependent fractality
are independent of whether we contemplate dilution or not (because
the height difference correlation function depends on strictly
local quantities~\cite{escuderojs}), and so we could, in this
particular calculation, substitute Eqs.~(\ref{hcubic2}) and
(\ref{hcubic2f}) by their dilution-free counterparts and still get
the same results. Note also that at the very beginning of this
section we have assumed the inequality $\zeta \le \delta < \zeta +
\gamma^{-1}$, which implies that for rapid growth the interface is
macroscopically but not microscopically uncorrelated. If $\delta
\ge \zeta + \gamma^{-1}$ then the interface is microscopically
uncorrelated and the fractal dimension becomes $d_f=d+1$
independently of the scale from which we regard it, i. e., scale
dependent fractality is a genuine effect of macroscopic
decorrelation, which disappears for strong damping causing
microscopic decorrelation.

Note that scale dependent fractality does not appear in
non-growing domain systems as for long times saturation is
achieved and the fractal dimension becomes constant (assuming no
multifractality is present). Although the behavior of the height
difference correlation function we found here is similar to the
one present in classical unbounded systems, results concerning the
fractal dimension cannot be immediately extrapolated. The fractal
dimension can be computed in a bounded growing domain, for
instance using the box counting method as we have done herein, by
employing as the reference length $L(t)$, the linear time
dependent size of the system. Of course, in an unbounded static
domain there is not such a reference length.

\section{The Kardar-Parisi-Zhang Equation}
\label{kpz}

One of the most important nonlinear models in the field of surface
growth is the Kardar-Parisi-Zhang (KPZ) equation~\cite{kpz}
\begin{equation}
\partial_t h= \nu \nabla^2 h + \lambda (\nabla h)^2 + \xi(x,t).
\end{equation}
It is related to the biologically motivated Eden model, as this
model, at least in a planar geometry, was numerically found to
belong to the KPZ universality class~\cite{barabasi}. As we will
see, understanding the KPZ equation on a growing domain may shed
some light on some of the properties of the classical version of
this model.

The KPZ equation on a growing domain reads
\begin{equation}
\label{dkpz}
\partial_t h= \nu \left( \frac{t_0}{t} \right)^{2 \gamma} \nabla^2 h +
\frac{\lambda}{2} \left( \frac{t_0}{t} \right)^{2 \gamma} (\nabla
h)^2 -\frac{d \gamma}{t}h + \gamma Ft^{\gamma-1} + \left(
\frac{t_0}{t} \right)^{d \gamma/2} \xi(x,t).
\end{equation}
Of course, if we just considered the dilatation $x \to
(t/t_0)^\gamma x$ we would find
\begin{equation}
\label{ndkpz}
\partial_t h= \nu \left( \frac{t_0}{t} \right)^{2 \gamma} \nabla^2 h +
\frac{\lambda}{2} \left( \frac{t_0}{t} \right)^{2 \gamma} (\nabla
h)^2 + \gamma Ft^{\gamma-1} + \left( \frac{t_0}{t} \right)^{d
\gamma/2} \xi(x,t).
\end{equation}
As we have shown in the previous section, the dilution mechanism
fixes the Family-Vicsek scaling in the fast growth regime. In the
radial Eden model case, assuming it belongs to the KPZ
universality class, we would have $z=3/2$ in $d=1$ and $\gamma=1$.
And so, one would naively expect that the resulting interface is
uncorrelated and we have to resort on dilution effects in order to
fix the Family-Vicsek ansatz and get rid of memory effects. But
here comes the paradoxical situation. There are two main
symmetries associated with the $d$-dimensional KPZ equation: the
Hopf-Cole transformation which maps it onto the noisy diffusion
equation~\cite{wio} and the related directed polymer
problem~\cite{kardar,lassig}, and Galilean invariance which have
been traditionally related to the non-renormalization of the KPZ
vertex at an arbitrary order in the perturbation
expansion~\cite{fns,medina}. In the case of the no-dilution KPZ
equation (\ref{ndkpz}) both symmetries are still present. Indeed,
this equation transforms under the Hopf-Cole transformation
$u=\exp[\lambda h/(2\nu)]$ to
\begin{equation}
\partial_t u = \nu \left( \frac{t_0}{t} \right)^{2 \gamma} \nabla^2 u +
\frac{\gamma F \lambda}{2 \nu} t^{\gamma-1} u + \frac{\lambda}{2
\nu} \left( \frac{t_0}{t} \right)^{d \gamma/2} \xi(x,t) u,
\end{equation}
which is again a noisy diffusion equation and it can be explicitly
solved in the deterministic limit $\epsilon=0$. We find in this
case
\begin{equation}
u(x,t)=\frac{(1-2\gamma)^{d/2} \exp[F \lambda t^\gamma/(2
\nu)]}{[4 \pi t_0^{2\gamma}(t^{1-2\gamma}-t_0^{1-2\gamma})]^{d/2}}
\int_{\mathbb{R}^d} \exp \left[ -\frac{|x-y|^2(1-2\gamma)}{4
t_0^{2 \gamma}(t^{1-2\gamma}-t_0^{1-2\gamma})} \right] u(y,t_0)
dy,
\end{equation}
which corresponds to
\begin{equation}
h(x,t)= \frac{2 \nu}{\lambda} \ln \left \{ \frac{(1-2\gamma)^{d/2}
\exp[F \lambda t^\gamma/(2 \nu)]}{[4 \pi
t_0^{2\gamma}(t^{1-2\gamma}-t_0^{1-2\gamma})]^{d/2}}
\int_{\mathbb{R}^d} \exp \left[ -\frac{|x-y|^2(1-2\gamma)}{4
t_0^{2 \gamma}(t^{1-2\gamma}-t_0^{1-2\gamma})} +\frac{\lambda}{2
\nu} h(y,t_0) \right] dy \right\},
\end{equation}
for given initial conditions $u(x,t_0)$ and $h(x,t_0)$. Note we
are using the same shorthand notation for differentials and
coordinates as in the Introduction. It is clear by regarding this
formula that decorrelation at the deterministic level will happen
for $\gamma > 1/2$. It is still necessary to find out if at the
stochastic level this threshold will be moved to $\gamma > 2/3$.
If we consider the dilution KPZ equation (\ref{dkpz}) then
transforming Hopf-Cole we would find the nonlinear equation
\begin{equation}
\partial_t u = \nu \left( \frac{t_0}{t} \right)^{2 \gamma} \nabla^2
u -\frac{d \gamma}{t} u \ln(u) + \frac{\gamma F \lambda}{2 \nu}
t^{\gamma-1} u + \frac{\lambda}{2 \nu} \left( \frac{t_0}{t}
\right)^{d \gamma/2} \xi(x,t) u,
\end{equation}
which may be thought of as a time dependent and spatially
distributed version of the Gompertz differential
equation~\cite{gompertz}. In this case it is not evident how to
find an explicit solution at the deterministic level and what
would be its decorrelation threshold.

Galilean invariance means that the transformation
\begin{equation}
x \to x-\lambda v t, \qquad h \to h+vx, \qquad F \to F -
\frac{\lambda}{2}v^2,
\end{equation}
where $v$ is an arbitrary constant vector field, leaves the KPZ
equation invariant. In case of no dilution this transformation can
be replaced by
\begin{equation}
x \to x-\frac{\lambda}{1-2\gamma} v t_0^{2 \gamma} t^{1-2\gamma},
\qquad h \to h+vx, \qquad F \to F - \frac{\lambda}{2 \gamma}v^2
t_0^{2 \gamma} t^{1-3\gamma},
\end{equation}
which leaves invariant equation~(\ref{ndkpz}). If we consider
dilution, then it is not clear how to extend this transformation
to leave equation~(\ref{dkpz}) invariant. The main difficulty
comes from the dilution term which yields a non-homogeneous
contribution to the dynamics as a response to the rotation $h \to
h+vx$. So in summary we may talk of a certain sort of Galilean
invariance which is obeyed by the no-dilution KPZ dynamics
(\ref{ndkpz}) and is lost when dilution is taken into account. If
it were found that the dilution equation~(\ref{dkpz}) obeys the
traditional KPZ scaling (at least in some suitable limit), then
that would mean the possible necessity for readdressing the role
that the symmetries of the KPZ equation have in fixing the
universality class~\cite{hochberg1,hochberg2,wio2,wio3,wio4}.

There is still another fundamental symmetry of the KPZ equation,
but this time it just manifests itself in one spatial dimension:
the so called fluctuation-dissipation theorem~\cite{barabasi}. It
basically says that for long times, when saturation has already
being achieved, the nonlinearity ceases to be operative and the
resulting interface profile would be statistically
indistinguishable from that created by the EW equation. For fast
domain growth, we know from the linear theory that the interface
never becomes correlated, and it operates, in this sense, as if it
were effectively in the short time regime for all
times~\cite{escuderojs}. As a consequence, the
fluctuation-dissipation theorem is not expected to play any role
in this case. Of course, this result would be independent of
whether we contemplated dilution or not.

In more general terms, it is known that the different symmetries
of statistical mechanical models influence their scaling
properties~\cite{henkel,pleimling}. It would be interesting to
understand in complete generality the interplay among the
symmetries of a physical model in a static domain and the
asymmetric presence of dilution when we let this domain grow in
time. A possible framework to carry out this project is the
instanton approach for the KPZ equation developed
in~\cite{fogedby1,fogedby2}. Complimentarily, this approach
motivates an interesting theoretical problem which is the
development of appropriate techniques to deal with an explicitly
time dependent Martin-Siggia-Rose theory, this is, with a
non-autonomous infinite dimensional Hamiltonian dynamical system.

\section{Center of mass fluctuations}
\label{comfluct}

Another property that has been studied in the context of radial
growth, particularly in Eden clusters, is the center of mass
fluctuations. In this section we derive for the first time the
properties of the center of mass fluctuations of the cluster
interfaces described by radial stochastic growth equations. It was
found numerically that the Eden center of mass fluctuates
according to the power law $C_m \sim t^{2/5}$ in
$d=1+1$~\cite{ferreira}, while in $d=2+1$ there is a strong
decrease in this exponent~\cite{madison}. This reduced stochastic
behavior in higher dimensions was already predicted in
\cite{escudero} using radial growth equations, and we will further
examine herein the compatibility among the equations and the Eden
cluster dynamics. The center of mass fluctuations are
characteristic not only of radial growth but also of planar
situations. Let us recall the classical EW equation
\begin{equation}
\partial_t h = D \nabla^2 h + \xi(x,t),
\end{equation}
defined on a one dimensional domain of linear size $L_0$ and with
no flux boundary conditions. It is straightforward to find that
the center of mass $h_0(t) = L_0^{-1}\int_0^{L_0} h(x,t)dx$ is a
Gaussian stochastic process defined by its two first moments
\begin{equation}
\left< h_0(t) \right>=0, \qquad \left< h_0(t) h_0(s) \right>=
\frac{\epsilon}{L_0} \min(t,s),
\end{equation}
and so we have found that the center of mass performs Brownian
motion, or equivalently we would say that its position is given by
a Wiener process. Note that the fluctuations amplitude decreases
with the linear system size, suggesting that in the case of a
growing domain our current law $C_m=\left< h_0^2 \right>^{1/2}
\sim t^{1/2}$ will be replaced by a different power law with a
smaller exponent. It is easy to see that this result does not hold
uniquely for the one dimensional EW equation; indeed, for any
$d-$dimensional growth equation with a conserved growth mechanism,
be it linear as the EW or Mullins-Herring
equations~\cite{barabasi} or nonlinear as the Villain-Lai-Das
Sarma equation~\cite{villain,lai} or its Monge-Amp\`{e}re
variation~\cite{escudero3}, the center of mass performs Brownian
motion characterized by the correlators
\begin{equation}
\left< h_0(t) \right>=0, \qquad \left< h_0(t) h_0(s) \right>=
\frac{\epsilon}{L_0^d} \min(t,s),
\end{equation}
as a consequence of the decoupling of the zeroth mode with respect
to the surface fluctuations~\cite{villain}. Note that in the case
of non-conserved growth dynamics this is not the case, as
illustrated by the KPZ equation
\begin{equation}
\partial_t h= \nu \nabla^2 h + \lambda (\nabla h)^2 + \xi(x,t).
\end{equation}
It is easy to see that in this case
\begin{equation}
\frac{dh_0}{dt}= \frac{\lambda}{L^d} \int (\nabla h)^2 dx
+\xi_0(t) \ge \xi_0(t),
\end{equation}
where $\xi_0(t)=L^{-d}\int \xi(x,t) dx$ and the equal sign is
attained only for $h= \,$constant, an unstable configuration for
KPZ dynamics. And so one expects stronger center of mass
fluctuations in this case. Actually, the short time center of mass
fluctuations can be easily calculated for any model which obeys
the Family-Vicsek scaling, including the KPZ equation. Indeed, the
Family-Vicsek scaling implies the following form of the
height-height correlation
\begin{equation}
\left< h(x,t)h(x',t) \right>= t^{2 \beta} \, C \left(
\frac{|x-x'|}{t^{1/z}} \right),
\end{equation}
which in the short time limit reduces to
\begin{equation}
\left< h(x,t)h(x',t) \right> \sim t^{2 \beta + d/z} \, \delta
\left( x-x' \right),
\end{equation}
leading to the result
\begin{equation}
\label{com}
\langle h_0(t)^2 \rangle \sim L^{-d} t^{2 \beta + d/z}.
\end{equation}
And so, within the Family-Vicsek scaling framework, the exponent
characterizing the short time behavior of the center of mass
fluctuations is $\beta +d/(2z)$.

As we have seen, the center of mass fluctuations are given by the
zeroth mode. In the growing domain case it can be shown that the
equation controlling the evolution of $h_0$ is \cite{escuderojs}
\begin{equation}
\frac{d h_0}{dt}= -\frac{d \gamma}{t}h_0+ \gamma F t^{\gamma-1} +
\left(\frac{t_0}{t}\right)^{d \gamma/2}\xi_0(t),
\end{equation}
in case dilution is taken into account. In this case we find for
long times the center of mass fluctuations
\begin{equation}
\label{comfv}
C_m^2= \left< h_0(t)^2 \right>-\left< h_0(t)
\right>^2=\frac{\epsilon t_0^{d \gamma}}{L_0^d (d \gamma +1)}
t^{1-d \gamma},
\end{equation}
and so $C_m \sim t^{(1-d \gamma)/2}$. If we did not consider
dilution we would find in the long time limit
\begin{equation}
\label{comfvnot} C_m^2 = \left\{
\begin{array}{lll} \frac{\epsilon \, t_0^{d \gamma}}{L_0^d (1-d \gamma)}t^{1-d \gamma} & \mbox{\qquad if \qquad $\gamma <
1/d$}, \\ \frac{\epsilon \, t_0}{L^d_0} \ln(t) & \mbox{\qquad if
\qquad $\gamma = 1/d$}, \\
\frac{\epsilon \, t_0}{L_0^d (d \gamma -1)} & \mbox{\qquad if
\qquad $\gamma > 1/d$}.
\end{array} \right.
\end{equation}
If we adapt result (\ref{com}) to the present setting we find
\begin{equation}
\left< h(x,t)h(x',t) \right> \sim t^{2 \beta + d/z} \, \delta
\left[ t^\gamma (x-x') \right] = t^{2 \beta + d/z-\gamma d} \,
\delta \left( x-x' \right).
\end{equation}
For linear systems the equality $2 \beta + d/z = 1$ holds, and so
this last equation agrees with (\ref{comfv}) but not with
(\ref{comfvnot}). This is a consequence of the violation of the
Family-Vicsek scaling in the absence of
dilution~\cite{escuderojs,escuderoar}. In the case of the
$(1+1)-$dimensional Eden model $d=\gamma=1$, and if it belonged to
the KPZ universality class the center of mass would fluctuate
according to the law $C_m \sim t^{1/6}$. This of course does not
agree with the measured behavior $C_m \sim t^{2/5}$. This exponent
could be recovered by introducing an \emph{ad hoc} instability
mechanism, such as for instance considering a growth equation
whose zeroth moment obeyed
\begin{equation}
\frac{d h_0}{dt}=D \left( \frac{t_0}{t} \right)^{\delta \gamma}
h_0 + \gamma F t^{\gamma-1}+ \left( \frac{t_0}{t} \right)^{d
\gamma/2} \xi_0(t).
\end{equation}
The desired exponent is obtained for $\delta=1$ and $Dt_0=2/5$,
but however this result is uniform on the spatial dimension and so
can not predict the $(2+1)-$dimensional behavior~\cite{madison}.
Additionally this instability mechanism seems to be not enough
justified and too non-generic to be a good explanation of the
observed phenomenology. Everything points to the fact that the
center of mass fluctuations of the Eden model result from a strong
violation of the Family-Vicsek scaling. As we may see from
equation (\ref{comfvnot}), this sort of violations imply stronger
center of mass fluctuations. This point will be further discussed
in the next section.

In summary we can say that the result $C_m \sim t^{2/5}$ suggests
a strong violation of the Family-Vicsek scaling by the surface
fluctuations of the $(1+1)-$dimensional Eden model. Although the
linear law $C_m \sim t^{(1-\gamma d)/2}$ does not reproduce
quantitatively the results, we still expect from it a qualitative
description of the dynamics, as the strong decrease of this
exponent was already reported in $(2+1)-$dimensions. According to
the linear law, the center of mass fluctuations should decrease
for increasing growth velocity and spatial dimension. Note also
that the nonlinearity seems to be a necessary ingredient; the
linearization of the KPZ equation proposed in~\cite{singha} reads
in Fourier space
\begin{equation}
\frac{d}{dt}\left< h_n^2 \right>=-A |n|^{3/2}\left< h_n^2
\right>+\frac{B}{|n|^{1/2}},
\end{equation}
for some constants $A$ and $B$ and in case of a non-growing
domain. This equation supports unbounded fluctuations as revealed
by the divergent stochastic contribution in the limit $n \to 0$,
and so this does not constitute a good model for predicting the
center of mass fluctuations.

\section{Applications to the Eden Model}
\label{eden}

In statistical mechanics it has been customary to classify the
behavior of discrete models within universality classes defined by
continuum field theories. Non-equilibrium growth theories have
been by no means an exception to this
rule~\cite{vvedensky1,vvedensky2}. In this sense one would be
interested in finding the universality class the Eden model
belongs to. According to the simulations performed in the planar
geometry the Eden model belongs to the KPZ universality
class~\cite{barabasi}. This agrees with the measured exponent
$\beta=1/3$ in radial systems~\cite{ferreira}. However, as we have
already seen, there are at least two possible universality classes
associated to the KPZ equation in radial systems: dilution-KPZ and
dilatation-KPZ. The first one is characterized by a behavior more
akin to that of planar systems, and the second one by memory
effects which imply the departure from the Family-Vicsek scaling.
According to the measurement of the autocorrelation exponent of
the Eden model in~\cite{singha} that yielded $\lambda=1/3$, {\it
the Eden model would be in the dilatation-KPZ universality class}
(one would expect $\lambda=4/3$ for dilution dynamics according to
the theory developed herein and in~\cite{escuderojs}). This fact
admits a simple explanation. In the Eden model, cells are
aggregated to the colony peripheral in such a way that the
positions of already present cells are not modified. Consequently,
as the system grows, no dilution is redistributing its
constituents. So the rigidity of the Eden model may well be at the
origin of the memory effects present at its
interface~\cite{singha}, which presumably place it in the
dilatation-KPZ universality class. But to be sure one would still
need, of course, to verify that this implies no contradiction with
the center of mass fluctuations as discussed in the last section.

As we have already mentioned, the Eden model may be thought of as
an idealization of a developing cell colony. Of course, as it was
completely clear from the very beginning~\cite{eden1,eden2}, there
are multiple factors of biological, chemical and even physical
nature that are not captured by this model. Apart from them, one
could be interested in improving the model in pure statistical
mechanical terms. To this end one may look for inspiration in real
cell colonies. The structure of a rapidly developing cell colony
would be dominated by dilution effects, originated in the birth of
new cells which volume causes the displacement of the existent
cells. This feature is not captured by any sort of Eden model
(diverse proliferation rules, on/off lattice,...) and is
fundamental in preserving the Family-Vicsek scaling, as we have
already seen. So it seems quite reasonable to modify the Eden
model in order to remove its rigidity, allowing bulk cells
proliferation and the displacement of the existent cells, both at
the bulk and interface, by the newborn cells. This would not be
interesting just in modelling terms, but also for introducing
dilution in the model and consequently shifting its universality
class.

\section{Conclusions and outlook}
\label{outlook}

In this work we have investigated the role of dilution and
decorrelation on radial growth. Dilution drives matter
redistribution along the growing interface: as the surface becomes
larger the already deposited matter occupies a smaller fraction of
interface, which is being simultaneously complemented with
incoming matter, the actual driving force of domain growth in
radial systems. Dilution is important for any rate of domain
growth, as it keeps the interfacial density constant, but
specially for rapidly growing domains, for which the diffusion
mechanism becomes irrelevant and dilution becomes the sole
responsible for the propagation of correlations on the macroscopic
scale. The importance of dilution is such that in its absence,
which takes place in the alternative dilatation dynamics, strong
memory effects arise. These include an enhanced stochasticity,
which separates the behavior of the large spatial scale limit of
the two-points correlation function from that dictated by the
Family-Vicsek scaling, and the appearance of non-universal
critical exponents in the marginally rough regime, characterized
by the equality $\zeta=d$. As have seen, both universality and the
Family-Vicsek structure of the correlation function are recovered
by virtue of dilution. This is at least what happens in the cases
mostly considered herein, which focus on unveiling the effects of
domain growth. Topological effects are indeed present when one
considers global scaling properties of hyperspherical
interfaces~\cite{escuderolast}.

As dilution propagates correlations at the same speed at which the
interface grows a global correlation becomes impossible for fast
domain growth. This leads to decorrelation, or in other words, to
a whitening of the interfacial profile in the sense that distant
points become uncorrelated. Decorrelation might be macroscopic,
which is evident only if we regard the dynamics from a spatial
scale of the same order of magnitude of the system size, or
microscopic, in which case it is apparent for much smaller length
scales. Microscopic decorrelation supports rapid roughening, i.
e., growth regimes characterized by $\beta >1/2$. These appear
naturally in the context of radial growth, for instance by
considering the IS equation, which results from a geometric
variational principle and for which $\zeta=d=2$ and $\delta=3$,
and thus it shows rapid roughening for all $\gamma > 1$. A
consequence of macroscopic decorrelation is the advent of a scale
dependent interfacial fractal dimension (so the surface becomes a
scale dependent fractal) which we have conjectured to be
self-similar.

There are several theoretical problems that can be
straightforwardly analyzed with the techniques introduced here. We
have for instance considered radial interfaces whose mean radius
grows as a power law of time $\left< r \right> \sim t^\gamma$.
This result has been obtained by means of a linear mechanism in
which an explicit power law dependence on time has been
considered, see Eq. (\ref{rdeposition}). This linear mechanism can
be substituted by a nonlinear one in which time does not appear
explicitly
\begin{equation}
\partial_t r = \gamma F^{1/\gamma} r^{1-1/\gamma}
+\frac{1}{r^{d/2}J(\vec{\theta})^{1/2}}\xi(\theta,t),
\end{equation}
which yields at the deterministic order $R=F t^\gamma$ again, but
it is the source at the first stochastic order of a term
(reminiscent of dilution) which may be either stabilizing or
destabilizing depending on the value of $\gamma$
\begin{equation}
\partial_t \rho = \frac{\gamma-1}{t} \rho
+\frac{1}{F^{d/2} t^{\gamma d/2}
J(\vec{\theta})^{1/2}}\eta(\theta,t);
\end{equation}
for small values of $\gamma$ the previous sections results are
recovered, while for large values of $\gamma$ memory effects and
enhanced (power law) stochasticity appear (which are standard
effects of instability as we have already seen), with the
threshold value of $\gamma$ depending of whether we introduce
dilution or not (in this concrete example dilution completely
erases instability). Also, this instability mechanism, contrary to
the ones studied herein and in~\cite{escudero2} which make the
zeroth mode unstable and the $l=1$ ones marginal, is able to
destabilize all modes. Different nonlinearities which might
destabilize a fixed number of modes lying before some given $l^*
\in \mathbb{N}$ can be easily devised too (basically by
introducing terms of the form $-r^{-m}$ for some suitable $m \in
\mathbb{N}$ in the corresponding equation of motion) and can even
be cast on some geometric variational formulation as the cases
considered in~\cite{escudero2}. Of course, deciding which model is
the good one must rely on numerical or experimental evidence based
on the study of specific models or systems of interest.

As mentioned in the introduction, part of the motivation for
studying radial growth models such as the Eden or different ones
lies in the possible similarity of these with some forms of
biological development, such as for instance cell colonies. The
results of our study can be translated into this context to obtain
some simple conclusions, provided the modelling assumptions make
sense for some biological system. The structure of a rapidly
developing cell colony would be dominated by dilution effects,
originated in the birth of new cells which volume causes the
displacement of the existent cells. If the rate of growth is large
enough this motion will dominate over any possible random
dispersal of the individual cells. It is remarkable that such a
consequence simply appears by considering domain growth, while it
is not necessary to introduce corrections coming from the finite
size of the constituents. This is the dilution dominated situation
we have formalized by means of the (decorrelation) inequality
$\gamma > 1/\zeta$ (assuming in this case $\delta=\zeta$). In this
case the overall appearance of the colony would be macroscopically
isotropic (this is, disregarding microscopic details). If we were
to introduce some control protocol in order to break this symmetry
we would need to eliminate colony constituents (possibly randomly
selected) at a high enough rate so the effective growth velocity
were one that reversed the decorrelation inequality. This would
make diffusion, instead of system size growth, the dominant
mechanism responsible for the colony macroscopic structure.
Consequently the macroscopic isotropy would be lost by means of
diffusion mediated anisotropic fluctuations developing on large
scales. This can be considered as a purely stochastic instability
which could perhaps be connected to the early stages of some
mechanism of biological pattern formation. For the one dimensional
Eden model, accepting it belongs to the KPZ universality class,
one finds $\gamma=1$ and $z =3/2$. If $z$ played the same role for
the nonlinear KPZ equation as $\zeta$ for the linear equations
considered herein (as it is reasonable to expect), the Eden model
would be in the uncorrelated regime. In order to control it we
would need to eliminate its cells at a rate such that the
effective growth rate obeyed $\gamma < 2/3$. For the two
dimensional Eden model, if its behavior were still analogous to
that of the KPZ equation, we would find $z > 3/2$ and thus a
greater difficulty for control. Note that for the particular
growth rules of the Eden model one would need to eliminate
peripheral cells in order to control the system. This would not be
so in the case of an actual bacterial colony, for which bulk cells
are still able to reproduce, and so cell elimination could be
performed randomly across the whole colony. Of course, these
conclusions are speculative as long as radial growth equations are
not proved to reasonably model some biological system.

In more general terms, we have found that the surface fluctuations
of the Eden model presumably strongly violate the Family-Vicsek
scaling. We have identified the absence of dilution in this model
as the reason underlying such a violation. In this sense, this
model would not be able to describe growing cell colonies,
precisely because it assumes a spurious rigidity of bulk cells. On
the other hand, it would be better suited to describe the radial
growth of crystalline structures~\cite{einstein}. We have also
found that reparametrization invariance as defined
in~\cite{maritan2} implicitly implies dilatation dynamics. Our
results call for an extension of the generalization of Langevin
dynamics to arbitrary geometries in order to capture both dilution
and dilatation scenarios, and the associated bifurcation of
universality classes. This same remark would affect as well
equilibrium systems, but in this case of course the domain
evolution will drive them out of equilibrium, unless growth is
quasistatic~\cite{parisi}.

\section*{Acknowledgments}

This work has been partially supported by the MICINN (Spain)
through Project No. MTM2010-18128.

\appendix

\section{Higher order perturbation expansion}
\label{horder}

As we have mentioned in Sec.~\ref{rrd}, the first order correction
in the small noise expansion is a Gaussian stochastic process. We
will try to go beyond this order in this appendix, and we will
show the difficulties that arise in trying so. We focus again on
the radial random deposition equation~(\ref{rdeposition}) and
assume the solution form
\begin{equation}
\label{snoise2} r(\vec{\theta},t)= R(t)+
\sqrt{\epsilon}\rho(\vec{\theta},t) + \epsilon
\rho_2(\vec{\theta},t),
\end{equation}
where the noise intensity $\epsilon$ will be used as the small
parameter~\cite{gardiner}. Substituting this solution form into
Eq.~(\ref{rdeposition}) we obtain the equations hierarchy
\begin{eqnarray}
\partial_t R &=& F \gamma t^{\gamma-1}, \\
\partial_t \rho_1 &=& \frac{1}{F^{d/2}t^{\gamma d/2}} \frac{\eta(\vec{\theta},t)}{J(\vec{\theta})^{1/2}},
\\
\partial_t \rho_2 &=& -\frac{d}{2 F^{1+d/2}} \frac{\rho_1}{t^{\gamma + d \gamma/2}}
\frac{\eta(\vec{\theta},t)}{J(\vec{\theta})},
\end{eqnarray}
where $\xi=\sqrt{\epsilon} \, \eta$ and both $\eta$ and $\xi$ are
now zero mean quasiwhite Gaussian processes whose correlations are
given by
\begin{equation}
\left< \eta(\vec{\theta},t) \eta(\vec{\theta},t) \right>=
C(\vec{\theta}-\vec{\theta}') \delta(t-t'), \qquad \left<
\xi(\vec{\theta},t) \xi(\vec{\theta},t) \right>= \epsilon
C(\vec{\theta}-\vec{\theta}') \delta(t-t'),
\end{equation}
where $C(\cdot)$ is some regular function approximating the Dirac
delta; the necessity for the quasiwhite assumption will we clear
in few lines. These equations have been derived assuming
$\sqrt{\epsilon} \ll F t^\gamma$, and we will further assume a
zero value for both initial perturbations as in Sec.~\ref{rrd}.
The solution to the first two was characterized in Sec.~\ref{rrd},
where the approximating function $C(\cdot)$ was substituted by the
Dirac delta. Here $R$ is a deterministic function and $\rho_1$ is
a zero mean Gaussian stochastic process that is completely
determined by its correlation function. The stochastic function
$\rho_2$ is a zero mean process too, but it is not Gaussian this
time, and its correlation (which no longer completely determines
the process) is given by
\begin{eqnarray}
\nonumber
\left< \rho_2(\vec{\theta},t) \rho_2(\vec{\theta}',s) \right> = \frac{d^2}{4F^{2+2d}(1-\gamma d)} \times \\
\left[ \frac{(\min\{t,s\})^{2-2\gamma-2\gamma
d}-t_0^{2-2\gamma-2\gamma d}}{2-2\gamma-2\gamma d} - t_0^{1-\gamma
d} \frac{(\min\{t,s\})^{1-2\gamma-\gamma d}-t_0^{1-2\gamma-\gamma
d}}{1-2\gamma-\gamma d} \right]
\frac{C(\vec{\theta}-\vec{\theta}')^2}{J(\vec{\theta})J(\vec{\theta}')},
\end{eqnarray}
if $\gamma d \neq 1$, $\gamma (1+d) \neq 1$, and $\gamma (2+d)
\neq 1$. If $\gamma d = 1$ we find
\begin{equation}
\left< \rho_2(\vec{\theta},t) \rho_2(\vec{\theta}',s) \right> =
\frac{1}{16 F^{2+2d}\gamma^4} \left\{t_0^{-2\gamma}- \left[
\min\{t,s\} \right]^{-2\gamma}\left[1+2\gamma
\mathrm{ln}\left(\frac{\min\{t,s\}}{t_0} \right)
\right]\right\}\frac{C(\vec{\theta}-\vec{\theta}')^2}{J(\vec{\theta})J(\vec{\theta}')},
\end{equation}
if $\gamma (1+d) = 1$ then
\begin{equation}
\left< \rho_2(\vec{\theta},t) \rho_2(\vec{\theta}',s) \right> =
\frac{d^2}{4F^{2+2d}\gamma}\left[ \mathrm{ln}\left( \frac{\min \{
t,s \}}{t_0} \right) + \frac{t_0^\gamma}{\gamma} ([\min \{ t,s
\}]^{-\gamma}-t_0^{-\gamma})
\right]\frac{C(\vec{\theta}-\vec{\theta}')^2}{J(\vec{\theta})J(\vec{\theta}')},
\end{equation}
and if $\gamma (2+d) = 1$ we get
\begin{equation}
\left< \rho_2(\vec{\theta},t) \rho_2(\vec{\theta}',s) \right> =
\frac{d^2}{8F^{2+2d}\gamma}\left[ \frac{(\min
\{t,s\})^{2\gamma}-t_0^{2\gamma}}{2\gamma}-t_0^{2\gamma}\mathrm{ln}\left(
\frac{\min \{t,s\}}{t_0} \right)
\right]\frac{C(\vec{\theta}-\vec{\theta}')^2}{J(\vec{\theta})J(\vec{\theta}')}.
\end{equation}
The long time behavior of the correlations, given by the condition
$t,s \gg t_0$, is specified by the following two-times and
one-time functions
\begin{eqnarray}
\left< \rho_2(\vec{\theta},t) \rho_2(\vec{\theta}',s) \right> &=&
\frac{d^2}{4F^{2+2d}(1-\gamma d)}
\frac{\left(\min\{ t,s \}\right)^{2-2\gamma-2\gamma d}}{2-2\gamma-2\gamma d} \frac{C(\vec{\theta}-\vec{\theta}')^2}{J(\vec{\theta})J(\vec{\theta}')}, \\
\left< \rho_2(\vec{\theta},t) \rho_2(\vec{\theta}',t) \right> &=&
\frac{d^2}{4F^{2+2d}(1-\gamma d)} \frac{t^{2-2\gamma-2\gamma
d}}{2-2\gamma-2\gamma d}
\frac{C(\vec{\theta}-\vec{\theta}')^2}{J(\vec{\theta})J(\vec{\theta}')},
\end{eqnarray}
when $\gamma (d+1)<1$, and if $\gamma (d+1)=1$ then
\begin{eqnarray}
\left< \rho_2(\vec{\theta},t) \rho_2(\vec{\theta}',s) \right> &=&
\frac{d^2}{4F^{2+2d}\gamma} \mathrm{ln}\left( \min\{t,s\} \right) \frac{C(\vec{\theta}-\vec{\theta}')^2}{J(\vec{\theta})J(\vec{\theta}')}, \\
\left< \rho_2(\vec{\theta},t) \rho_2(\vec{\theta}',t) \right> &=&
\frac{d^2}{4F^{2+2d}\gamma} \mathrm{ln}(t)
\frac{C(\vec{\theta}-\vec{\theta}')^2}{J(\vec{\theta})J(\vec{\theta}')},
\end{eqnarray}
and finally, when $\gamma (d+1)>1$, we find
\begin{equation}
\left< \rho_2(\vec{\theta},t) \rho_2(\vec{\theta}',s) \right> =
\frac{d^2}{8F^{2+2d}}\frac{t_0^{2-2\gamma-2\gamma
d}}{1-(3+2d)\gamma+
(2+3d+d^2)\gamma^2}\frac{C(\vec{\theta}-\vec{\theta}')^2}{J(\vec{\theta})J(\vec{\theta}')},
\end{equation}
a correlation function that vanishes in the limit $t_0 \to
\infty$. Now it is clear why we needed the quasiwhite
approximation: for a regular function $C(\cdot)$ the expression
$C(\cdot)^2$ makes sense, contrary to what happens if we
substitute it by the Dirac delta to get $\delta(\cdot)^2$. This is
the first indication of the failure of the higher order
perturbation theory.

We now examine the effect that dilution has on the random function
$\rho_2$, which in this case obeys the equation
\begin{equation}
\partial_t \rho_2=-\frac{\gamma d}{t} \rho_2 -\frac{d}{2}
\frac{(d+1)^{1+d/2}}{F^{1+d/2}t^{\gamma+\gamma
d/2}}\frac{\rho_1(\vec{\theta},t)\xi(\vec{\theta},t)}{J(\vec{\theta})}.
\end{equation}
In this case the long time correlation function reads
\begin{equation}
\left< \rho_2(\vec{\theta},t) \rho_2(\vec{\theta}',s) \right>=
\frac{d^2(d+1)^{2+2d}}{8F^{2+2d}(\gamma d +1)(1-\gamma)}
(ts)^{-\gamma d} \min\{t,s\}^{2-2\gamma}
\frac{C(\vec{\theta}-\vec{\theta}')^2}{J(\vec{\theta})J(\vec{\theta}')},
\end{equation}
if $\gamma <1$,
\begin{equation}
\left< \rho_2(\vec{\theta},t) \rho_2(\vec{\theta}',s) \right>=
\frac{d^2(d+1)^{1+2d}}{4F^{2+2d}} (ts)^{-d} \ln[\min\{t,s\}]
\frac{C(\vec{\theta}-\vec{\theta}')^2}{J(\vec{\theta})J(\vec{\theta}')},
\end{equation}
if $\gamma =1$,
\begin{equation}
\left< \rho_2(\vec{\theta},t) \rho_2(\vec{\theta}',s) \right>=
\frac{d^2(d+1)^{2+2d}}{8F^{2+2d}(\gamma d +1)(\gamma-1)}
(ts)^{-\gamma d} \, t_0^{2-2\gamma}
\frac{C(\vec{\theta}-\vec{\theta}')^2}{J(\vec{\theta})J(\vec{\theta}')},
\end{equation}
if $\gamma >1$. The one time correlation function is then
\begin{equation}
\left< \rho_2(\vec{\ell},t) \rho_2(\vec{\ell}',t) \right>=
\frac{d^2(d+1)^{2+2d}}{8F^{2+2d}(\gamma d +1)(1-\gamma)}
t^{2-2\gamma}
\frac{C(\vec{\ell}-\vec{\ell}')^2}{J(t^{-\gamma}\vec{\ell})J(t^{-\gamma}\vec{\ell}')},
\end{equation}
if $\gamma <1$,
\begin{equation}
\left< \rho_2(\vec{\ell},t) \rho_2(\vec{\ell}',t) \right>=
\frac{d^2(d+1)^{1+2d}}{4F^{2+2d}} \ln(t)
\frac{C(\vec{\ell}-\vec{\ell}')^2}{J(t^{-\gamma}\vec{\ell})J(t^{-\gamma}\vec{\ell}')},
\end{equation}
if $\gamma =1$,
\begin{equation}
\left< \rho_2(\vec{\ell},t) \rho_2(\vec{\ell}',t) \right>=
\frac{d^2(d+1)^{2+2d}}{8F^{2+2d}(\gamma d +1)(\gamma-1)}
t_0^{2-2\gamma}
\frac{C(\vec{\ell}-\vec{\ell}')^2}{J(t^{-\gamma}\vec{\ell})J(t^{-\gamma}\vec{\ell}')},
\end{equation}
if $\gamma>1$, where $\vec{\ell}-\vec{\ell}'=t^{\gamma}
(\vec{\theta}-\vec{\theta}')$,
$C(\vec{\ell}-\vec{\ell}')=t^{-\gamma d}
C(\vec{\theta}-\vec{\theta}')$, and we have assumed that the
approximating function $C(\cdot)$ has the same homogeneity as the
Dirac delta. Although it is evident that dilution carries out a
measurable action, particularly erasing part of the memory
effects, the result is far from satisfactory. In all cases the
prefactor deviates from the expected random deposition form
$t^2$~\cite{footnote}, the unexpected critical value $\gamma=1$
has appeared, and for $\gamma \ge 1$ memory effects are present as
signaled by the logarithm and the $t_0$ dependence respectively;
and the situation is further complicated by the presence of the
factor $C(\cdot)^2$ which becomes singular in the white noise
limit. All of these elements suggest the failure of the small
noise expansion beyond the first order. Classical results suggest
the possibility of constructing a systematic approach to the
solution of some nonlinear stochastic differential equations by
continuing the small noise expansion to higher
orders~\cite{gardiner}. Our present results suggest the failure of
this sort of expansions beyond the Gaussian (which turns out to be
the first) order in very much the same way as the Kramers-Moyal
expansion of the master equation~\cite{pawula} and the
Chapman-Enskog expansion of the Boltzmann equation~\cite{carlo}
fail beyond the Fokker-Planck and Navier-Stokes orders
respectively.


\begin{thebibliography} {99}
\bibitem{footnote} This prefactor always shows a smaller exponent,
and the expected $t^2$ only appears in the limit $\gamma \to 0$.
This reduction of the prefactor exponent happens equally in
presence and absence of dilution, and is the opposite trend to the
one observed in the dilution-free Gaussian order, which tends to
increase this exponent~\cite{escuderojs,escuderoar}. All this
strongly suggests the failure of the perturbation theory at this
order, rather than an effect related to dilution/dilatation.
\end{thebibliography}
\end{document}